\documentclass{article}

\usepackage{graphicx} 
\usepackage[margin=1in]{geometry} 

\usepackage{comment}
\usepackage{bbm}
\usepackage[dvipsnames, table]{xcolor}
\usepackage{mathtools} 
\usepackage{natbib}
\usepackage{graphicx} 
\usepackage{listings}
\usepackage{enumerate}
\usepackage{scalerel} 
\usepackage{amsmath}
\usepackage{amssymb}
\usepackage{amsfonts}
\usepackage{booktabs}
\usepackage{multirow}
\usepackage{threeparttable}
\usepackage{subcaption}

\usepackage{float}
\usepackage{caption}
\usepackage{colortbl}
\newcommand{\nothere}[1]{}

\usepackage{algorithm}
\usepackage{algorithmic}
\usepackage{orcidlink}
\usepackage{adjustbox}

\usepackage{verbatim}
\usepackage{tikz}
\usepackage{academicons}
\usetikzlibrary{shapes, positioning, arrows.meta, arrows}

\let\proglang=\textsf
\newcommand{\pkg}[1]{{\fontseries{b}\selectfont #1}}

\usepackage{fancyvrb}
\DefineVerbatimEnvironment{Sinput}{Verbatim}{fontshape=sl}
\DefineVerbatimEnvironment{Soutput}{Verbatim}{}
\DefineVerbatimEnvironment{Scode}{Verbatim}{fontshape=sl}

\DefineVerbatimEnvironment{Code}{Verbatim}{}
\DefineVerbatimEnvironment{CodeInput}{Verbatim}{fontshape=sl}
\DefineVerbatimEnvironment{CodeOutput}{Verbatim}{}
\newenvironment{CodeChunk}{}{}
\setkeys{Gin}{width=0.8\textwidth}

\title{\pkg{TimeDepFrail}: Time-Dependent Shared Frailty Cox Models in \proglang{R}}
\author{Alessandra Ragni$^{*}$\orcidlink{0000-0002-3647-7340}, Giulia Romani$^{*}$, Chiara Masci$^{*}$\orcidlink{0000-0002-9208-3194} \\
\small{$^*$MOX, Department of Mathematics, Politecnico di Milano,} \\
\small{Piazza Leonardo da Vinci 32, 20133, Milano, Italy}}
\date{}

\begin{document}

\maketitle

\abstract{
  This paper introduces \pkg{TimeDepFrail}, an \proglang{R} package designed to implement time-varying shared frailty models by extending the traditional shared frailty Cox model to allow the frailty term to evolve across time intervals. These models are particularly suited for survival analysis in clustered data where unobserved heterogeneity changes over time, providing greater flexibility in modeling time-to-event data.
  
  The package builds on the piecewise gamma frailty model originally proposed by Paik (1994) and refined by Wintrebert et al. (2004). Our key contributions include the integration of posterior frailty estimation, a reduction in computational complexity, the definition of a prediction framework and the efficient implementation of these models within an \proglang{R} package.
  As a practical application, we use \pkg{TimeDepFrail} to analyze dropout rates within a university, where high dropout rates are a known issue. By allowing frailty to vary over time, the package uncovers new insights into the unobserved factors influencing dropout. 
  \pkg{TimeDepFrail} simplifies access to advanced time-varying frailty models, providing a practical and scalable alternative to more computationally demanding methods, making it highly applicable for large-scale datasets.
}

\vspace{0.3cm}


\section{Introduction} \label{Introduction}

Survival analysis plays an important role in many fields, as it aims to model and analyze time-to-event data. One of the most widely used tools for this purpose is the \textit{Cox proportional hazards model} \citep{cox1972regression}, a semiparametric model that allows for the inclusion of both categorical and numerical covariates to predict the hazard function - the instantaneous risk of an event over time. While the Cox model captures \textit{observed heterogeneity} through relevant covariates \citep{balan2020tutorial}, it does not account for unmeasured factors influencing survival times, often referred to as \textit{unobserved heterogeneity}. This limitation can lead, in certain situations, to biased or incomplete interpretations of model outcomes.

To address unobserved heterogeneity, the concept of \textit{frailty} was introduced by \cite{vaupel1979impact}, representing a unit-level random effect that acts multiplicatively on the hazard function. Typically assumed to follow a gamma distribution, the frailty captures the influence of unmeasured factors on survival outcomes. In clustered data, where units belong to distinct groups, shared frailty models are commonly used, with a single frailty term shared among units within each group to model their dependent survival times \citep{balan2020tutorial, therneau2003penalized}. This \textit{Shared Frailty Cox Model} assumes that frailty remains constant throughout the follow-up period, which may be overly restrictive in many real-world applications.

In practice, group-specific characteristics often change over time. To address this, time-varying or dynamic frailty models have emerged, extending the classic frailty framework by allowing frailty to vary over time. These models provide a more flexible representation of survival data by accommodating time-varying unobserved heterogeneity within groups.

In particular, \cite{paik1994multivariate} and \cite{wintrebert2004centre} introduced a framework where the follow-up period is partitioned into a series of time intervals, with frailty terms allowed to vary across both groups and intervals. This transforms the survival model into a fully parametric one, wherein different baseline hazard components are estimated for each time interval. 
In a different context, \cite{paddy2012relative} and \cite{unkel2014time} developed flexible families of time-varying frailty models tailored for paired serological survey data.
Additionally, autoregressive frailty models have been explored to induce temporal correlations in the frailty process. For example, \cite{henderson2003serially} and \cite{fiocco2009new} proposed autocorrelated gamma frailty models, which were later extended to continuous time survival data by \cite{putter2015dynamic}. In their work, the frailty process for clustered survival data is constructed using compound birth-death processes in two dimensions. This approach differs from that of \cite{gjessing2003frailty}, who based their model on Lévy-type processes in one dimension for univariate survival data without clustering.
Autoregressive log-normal frailty models have also been introduced in the literature by \cite{mcgilchrist1996survival} and \cite{yau1998ml}. 
More recently, \cite{munda2016testing} proposed an approach where, instead of modeling the hazard function directly, the log cumulative hazard function is modeled. By incorporating time-varying random effects within a mixed model framework, this method allows for the testing of decreasing heterogeneity in clustered survival data over time.

In this context, the only available \proglang{R} package is \pkg{dynfrail} \citep{dynfrail}, which implements the model proposed by \cite{putter2015dynamic}. 
However, due to the complexity of the underlying model, the package suffers from slow computational performance, making it impractical for large datasets and limiting its feasibility for real data applications.

In this work, we present an \proglang{R} package \citep{timedepfrail2025} based on the time-varying frailty model proposed by \cite{wintrebert2004centre} and inspired by the work of \cite{paik1994multivariate}. Our contribution not only implements the proposed time-varying frailty model but also improves it by incorporating posterior frailty estimation, reducing the computational burden and making the model more suitable for large-scale datasets or complex applications. 
In addition, we incorporate a function for the estimation of the conditional survival function.
To illustrate the \pkg{TimeDepFrail} package, we conduct a case study with real-world data from an Italian university, analyzing dropout rates across different faculties. The model assesses the influence of time-varying unobserved factors on student dropout rates, illustrating the practical advantages of using a time-dependent shared frailty approach for analyzing grouped survival data.

The paper is organised as follows. Section \ref{statistical_background} provides the statistical background and theoretical concepts that will be used throughout. Section \ref{Methodology} details the employed methodology and functionality of the package, with corresponding pseudo-code for clarity.
In Section \ref{package}, we offer an overview of the package syntax and describe its key functions.
Section \ref{Application} presents a reproducible example based on university data, covering the package's complete functionality.
Finally, Section \ref{Conclusion} summarizes the features of \pkg{TimeDepFrail}, discussing its limitations and potential directions for future developments.

\section{Background and Concepts}
\label{statistical_background}

In this section, we introduce some useful notation in \textit{survival analysis} \citep{kleinbaum1996survival} and describe the main theory behind the \textit{Shared Frailty Cox models} \citep{cox1972regression, balan2020tutorial, therneau2003penalized} and the time-varying ones introduced by \cite{wintrebert2004centre, paik1994multivariate}.

\subsection{Basic notation in survival analysis}\label{basis_survival_analysis}

Let us consider a cohort of $n$ units, denoted by index $i \in \{1, \dots, n\}$ and let $T_i$ be the random \textit{time-to-event}.

Denoting by $t$ any specific value of interest of the non-negative random variable $T_i$, the \textit{survival function} $S_i(t) = P(T_i > t) = 1 - F_{T_i}(t)$ is defined as the probability for  unit $i$ to survive (i.e., not experiencing the event of interest) longer than time $t$, 
where $F_{T_i}(t)$ indicates the cumulative distribution function of $T_i$. 

The \textit{hazard function} $h_i(t) = \lim_{\Delta t \to 0} P(t \leq T_i \leq t + \Delta t \mid T_i \geq t) / \Delta t$ represents the instantaneous risk of experiencing the event of interest, conditionally on the fact that it has not occurred yet. 

Let now $C_i$ be the random \textit{right-censoring time}.
The outcome data in survival analysis consists of a couple of variables $(\tilde{T}_i, \delta_{i})$ for each $i$, where $\tilde{T}_i = min\{C_{i}, T_{i}\}$ and $\delta_{i} = 1 \text{ if } T_{i} \leq C_{i}$, 0 otherwise.

\subsection{Time-Invariant Shared Frailty Cox Model}\label{time_inv_shared_gamma_frailty_cox_model}

Let $i$, for $i=1,\ldots,n_j$ be a unit nested within group $j$, for $j=1\,\ldots,N$. In the \textit{(Time-Invariant) Shared Frailty Cox Model} \citep{cox1972regression, balan2020tutorial, therneau2003penalized}, 
the hazard function $h_{ij}(t)$ is modeled as: 
\begin{equation*}
    h_{ij}(t|\mathbf{x}_{ij}, Z_j) = Z_j \cdot h_0(t) \cdot \text{exp}(\boldsymbol{\beta}^T \boldsymbol{x}_{ij})
\end{equation*}
where $h_0(t)$ is the \textit{baseline hazard function}, $\mathbf{x}_{ij}$ is the R-dimensional column vector of $ij$-specific covariates, $\boldsymbol{\beta}$ is relative column vector of regression parameters ($^T$ stands for the transpose) and $Z_j$ is the frailty term shared by all units in group $j$, acting multiplicatively on $h_0(t)$ and accounting for \textit{unobserved heterogeneity}.
Conditionally on $Z_j$, the survival times of units within group $j$ are assumed to be independent.

Various prior frailty distributions can be used to model the unobserved heterogeneity. The most common is the \textit{Gamma} distribution, which is the basis for the \textit{Time-Independent Shared Gamma-Frailty Cox Model}.
Appendix \ref{app:Time-Independent Shared Gamma-Frailty Cox Model} contains further details, as well as the computation of the posterior mean and variance to which we will refer to in following sections.

\subsection{Time-Dependent Shared Frailty Cox model}
\label{adapted_paik_eam}

Among the various approaches proposed in literature for addressing the time-dependent shared frailty Cox models and recalled in the \textit{Introduction}, we here recall and focus on what is presented in \cite{wintrebert2004centre}. 

Let $t_{ij}$ be the time-to-event for unit $i$ in group $j$. 
The time-domain is divided into $L$ intervals $I_k = [a_{k-1},a_{k})$, $k \in \{1, \dots, L\}$, with discrete time-points $0=a_0 < a_1 < \dots < a_L = \infty$. In this way, $d_{ijk}$ is the event variable in $I_k$ for unit $i$ in $j$, such that $d_{ijk} = 1$ if $t_{ij}$ is uncensored in $I_{k}$ and 0 otherwise. The usual binary indicator $d_{ij} = \delta_{ij}$ of event can be obtained as $\sum_{k}{d_{ijk}}$.

Let $Z_{jk}$ be the unobservable frailty of group $j$, in time-interval $I_k$. Conditionally on $Z_{jk}$, the hazard function $h_{ijk}$ for $i$ in $j$ and $I_k$ is given by the general expression:
\begin{equation}\label{individual_frailty_dep_hazard}
    h_{ijk}(t_{ij} | Z_{jk}) = Z_{jk} \text{ exp}{(\boldsymbol{\beta}^T \boldsymbol{x}_{ij}  + \phi_k)} 
\end{equation}
where $\boldsymbol{x}_{ij}$ and $\boldsymbol{\beta}$ are, respectively, the usual R-dimensional unit vector of covariates and parameters, while $\phi_k$ is the newly introduced baseline log-hazard for interval $I_k$, $\forall k$.\\ 
Both the coefficients $\boldsymbol{\beta}$ and the baseline log-hazard need to be estimated with a suitable procedure. Their cardinality uniquely depends on the number of covariates $R$ and on the partition of the time-domain $L$. \\

The \textit{Adapted Paik et al.'s Model} (\textit{Adapted Paik eaM}) is one of the three models discussed in \cite{wintrebert2004centre}, inspired by \cite{paik1994multivariate}, which will be object of our focus\footnote{A complete but not efficient implementation of the other two models described in \cite{wintrebert2004centre} (the \textit{Centre-Specific Frailty Model with Power Parameter} and the \textit{Stochastic Time-Dependent Centre-Specific Frailty Model}) is available on \href{https://github.com/alessandragni/TimeDepFrail}{https://github.com/alessandragni/TimeDepFrail}.}.
The time-varying frailty term 
is defined as $Z_{jk}(t_{ij}) = (\alpha_j + \epsilon_{jk})$ for $t_{ij} \in I_k$, where 
\begin{itemize}
    \item $\alpha_j \sim Gamma(\mu_1/\nu, 1/\nu)$ $\forall j$ 
    \item $\epsilon_{jk} \sim Gamma(\mu_2/\gamma_k, 1/\gamma_k)$ $\forall j,k$
\end{itemize}
independent and such that $\mu_1, \mu_2, \nu, \gamma_k > 0$ $\forall k$ and the constraint $E[Z_{jk}] = \mu_1 + \mu_2 = 1$ needed for identifiability. 
The full list of the model parameters to be estimated is therefore composed of: $\phi_k$ $\forall k$, $\beta_r$ $\forall r, \mu_1, \nu$ and $\gamma_k \forall k$.\\\\
One important consideration about the frailty $Z_{jk}$ is that its expectation is identically equal to $1$ in each interval, but its variance is allowed to vary between different intervals. Moreover, thanks to the independence of $\alpha_j$ and $\epsilon_{jk}$, $\forall j,k$, this variance can be derived as the sum of the variance of its components: $var(Z_{jk}) = \mu_1 \nu + \mu_2 \gamma_k$.
By looking at the structure of the frailty, we observe that only $\epsilon_{jk}$ models the time-dependence effect of each group, while $\alpha_j$ represents the constant effect associated to each one of them. Consequently, also the variance just defined is given by the sum of two different contributes: a time-varying one ($\mu_2 \gamma_k$) and a constant one ($\mu_1 \nu$), that only moves upward the entire function. Thus, an effective representation of the variance could be obtained by the sole time-varying term $\mu_2 \gamma_k$. \\\\
The full log-likelihood has the form (see Appendix~\ref{proof_loglikleihood_function} for further details):
\begin{align}
        l &= \sum_{j=1}^{N} \left[ \sum_{i,k} d_{ijk}(\boldsymbol{\beta}^T \boldsymbol{x}_{ij}+\phi_k) - \frac{\mu_1}{\nu}\text{log}(1+\nu A_{j..}) + \sum_{k} \left[ \frac{-\mu_2}{\gamma_k} \text{log}(1+\gamma_kA_{j.k})\right]\right] + \nonumber\\
        &+ \sum_{j=1}^{N} \left[ \sum_{k} \left[ \text{log}\left( \sum_{l=0}^{d_{j.k}}\binom{d_{j.k}}{l} \frac{\Gamma(\mu_2/\gamma_k+d_{j.k}-l)}{\Gamma(\mu_2/\gamma_k)} \frac{\Gamma(\mu_1/\nu+l)}{\Gamma(\mu_1/\nu)} \frac{(A_{j.k}+1/\gamma_k)^{(l-d_{j.k})}}{(A_{j..}+1/\nu)^l} \right) \right] \right] \label{ll_AdPaik}
\end{align}

where $A_{ijk} = e_{ijk}\text{e}^{(\boldsymbol{\beta}^T \boldsymbol{x}_{ij} + \phi_k)}$, $A_{j.k} = \sum_{i}A_{ijk}$, $A_{j..} = \sum_{i,k} A_{ijk}$, $d_{j.k}=\sum_{i}d_{ijk}$ and
\begin{equation}
    e_{ijk} =\begin{cases}  0 & \text{if } t_{ij} < a_{k-1} \\ t_{ij}-a_{k-1} & \text{if }t_{ij} \in I_k \\ a_{k}-a_{k-1} & \text{if }t_{ij} \geq a_k 
\end{cases}
\end{equation}
%

\subsubsection{Posterior frailty estimates}
The setting presented in \cite{wintrebert2004centre} lacks a posterior inference procedure. We therefore propose here a method to 
compute \textit{a posteriori} the frailty terms, given the data and the estimated parameters. Since the components of the frailty are independent and both distributed according to a Gamma, our proposal is to replicate the procedure of the time-invariant shared gamma-frailty model recalled in Appendix \ref{app:Time-Independent Shared Gamma-Frailty Cox Model} and to derive separate estimates for each term, group and, possibly, interval of the time-domain. For further details about how these quantities are derived and computed, see Appendix~\ref{proof_posterior_frailty_estimates_posterior_frailty_variance}. By this approach, we get:
\begin{align} 
    \hat{\alpha}_j = \frac{(\hat{\mu}_1/\hat{\nu}) + N_j}{(1/\hat{\nu}) + \hat{H}_{j,\bullet}}& \quad \quad 
    \hat{\epsilon}_{jk} = \frac{(\hat{\mu}_2/\hat{\gamma_k}) + N_j(I_k)}{(1/\hat{\gamma_k}) + \hat{H}_{j,\bullet}(I_k)}  \quad \quad    \hat{Z}_{jk} = \frac{\hat{\alpha}_j}{\hat{\alpha}_{max}} + \frac{ \hat{\epsilon}_{jk}}{\hat{\epsilon}_{max}} \quad \forall j,k \label{post_frailty_estimates}\\ \\
    &\hat{\alpha}_{max} = \max_j \hat{\alpha}_j \quad \quad \quad \quad \hat{\epsilon}_{max} = \max_{j,k} \hat{\epsilon}_{jk} \nonumber
\end{align}
being $N_j(I_k)$ and $\hat{H}_{j,\bullet}(I_k)$, respectively, the number of events and cumulative hazard function in group $j$ and interval $I_k$, while $N_j$ and $\hat{H}_{j,\bullet}$ are evaluated at the end of the follow-up, so that

\begin{align}
    & var(\hat{\alpha}_j/\hat{\alpha}_{max}) = \frac{(\hat{\mu}_1/\hat{\nu}) + N_j}{((1/\hat{\nu}) + \hat{H}_{j,\bullet})^2} \cdot \frac{1}{(\alpha_{max})^2} \nonumber\\
    & var(\hat{\epsilon}_{jk}/\hat{\epsilon}_{max}) = \frac{(\hat{\mu}_2/\hat{\gamma_k}) + N_j(I_k)}{((1/\hat{\gamma_k}) + \hat{H}_{j,\bullet}(I_k))^2} \cdot \frac{1}{(\epsilon_{max})^2} \label{post_frailty_variance}\\
    & var(\hat{Z}_{jk}) = var(\hat{\alpha}_j/\hat{\alpha}_{max}) + var(\hat{\epsilon}_{jk}/\hat{\epsilon}_{max}) \quad \quad  \forall j,k. \nonumber
\end{align}
For what concerns the asymptotically normal $95\%$ confidence intervals for the posterior frailty $Z_{jk}$, we employ the variance of the posterior frailty distribution, so that
\begin{equation}\label{ci_post_frailty_estimates}
    CI_{0.95}(Z_{jk}) = [\hat{Z}_{jk} \pm 1.96 \cdot \sqrt{var(\hat{Z}_{jk})}].
\end{equation}

\subsubsection{Parameter estimation}
To estimate all the parameters of the \textit{Adapted Paik et al.'s Model} we maximize the log-likelihood function in (\ref{ll_AdPaik}) using a suitable optimization method, corresponding to a reintepretation of the \textit{Powell's method} \citep{powell1964efficient} in a multidimensional setting. For each parameter $p$, we compute the point estimate $\hat{p}$ and the $95\%$ confidence interval (default) as
\begin{equation}\label{ci_params}
    CI(p) = [\hat{p} \pm 1.96 \cdot se(\hat{p})]
\end{equation}
The standard error $se(\hat{p})$ for all the estimated parameters is computed as the inverse of the square root of the \textit{Information matrix}, evaluated as the opposite of the \textit{hessian matrix} of the log-likelihood function. An accurate description of the optimization procedure, the computation of the parameter standard error and the numerical approximation of the hessian matrix is contained in the next Section. 

\subsubsection{Conditional survival function}
The setting presented in \citet{wintrebert2004centre} also lacks a procedure for the computation of the conditional survival function.
We therefore adapt the general expression, that follows from \citet{putter2015dynamic} and that assumes the form $S_{ij}(t \mid Z_j) = \exp \big\{ - \int_0^t Z_j(s) h_0(s) e^{\boldsymbol{\beta}^T \mathbf{x}_{ij}} ds \big\}$, to our case as follows:
\begin{equation}
    \hat{S}_{ij}(t_{ij} \mid \hat{Z}_{jk}) = \exp \bigg\{ - e^{\hat{\boldsymbol{\beta}}^T \boldsymbol{x}_{ij}} \cdot \sum_k \hat{Z}_{jk} \cdot \exp(\hat{\phi}_k) \cdot \Delta_{I_k}\bigg\}
    \label{eq:condsurvfun}
\end{equation}
being $\Delta_{I_k}$ the length of each interval $I_k$.

\section{Setup and Methods}\label{Methodology}
This present section is organised as follows. In Subsection \ref{setup_variables} we describe the variables setup, follows the computation of model log-likelihood in Subsection \ref{computation_loglik}, the algorithm for maximizing it in Subsection \ref{max_like} and computation of other outputs in Subsection \ref{outputs}.

\subsection{Setup variables}\label{setup_variables}
\subsubsection{Temporal variables}

Following \citet{wintrebert2004centre}, first of all, the time-domain needs to be divided into intervals of possibly varying lengths, with the right boundary fixed at the study's end and the left boundary at the start of follow-up or shortly after, depending on when events begin. Denote with $T_{\text{axis}}$ the vector of the partitioned time-domain (e.g., $T_{\text{axis}} = [1.0, 2.0, 3.0]$).
The internal subdivision is data-specific, and guidelines on how to proceed are discussed in Subsection~\ref{inputs}. 

For the time-to-event variable $t_{ij}$ introduced in Section~\ref{adapted_paik_eam}, representing when unit $i$ in group $j$ experiences the event, we assign a default value to units who do not face the event during follow-up. This value is set just past the end of the time-domain (e.g., if follow-up ends at $t=6$, $t_{ij}$ could be $6.1$ or greater) to facilitate integral computations for the likelihood (see Appendix~\ref{proof_loglikleihood_function}).

Finally, we define the temporal event variable $d_{ijk}$ as the time-varying extension of the dropout variable $d_{ij}$. Both $d_{ijk}$ and the event indicator $e_{ijk}$ are constructed as matrices, with rows representing units and columns representing time intervals.

\subsubsection{Vector of parameters}
The model parameters can be distinguished into three main types: the Cox regression coefficients $\boldsymbol{\beta} = [\beta_1, \dots, \beta_R]$, the baseline log-hazards $\boldsymbol{\phi} = [\phi_1,\dots,\phi_L]$, and time-varying frailty terms $\mu_1$, $\nu$, $\boldsymbol{\gamma} = [\gamma_1, \dots, \gamma_L]$ composing $Z_{jk}$, where $\mu_1, \nu, \gamma_1, \dots, \gamma_L > 0$. These parameters are collectively represented as:
\begin{align}
    &\boldsymbol{p} = [\boldsymbol{\phi}, \boldsymbol{\beta}, \mu_1, \nu, \boldsymbol{\gamma}] \label{parameters_disposition} 
\end{align}
We also define the corresponding numerosity vector $\boldsymbol{n_p} = [L, R, 1, 1, L]$, so that $\sum \boldsymbol{n_p} = 2 \cdot L + R + 2$ indicates the total number of parameters.

To estimate the parameters in $\boldsymbol{p}$, as will be discussed in the following sections, we use a constrained optimization procedure to maximize the likelihood function in (\ref{ll_AdPaik}).
Some parameters require essential constraints to ensure valid estimates (e.g., positivity of certain parameters). Other constraints, while not strictly necessary, are imposed in order to improve the efficiency of the optimization process.
In both cases, reasonable parameter ranges should be assigned for randomly selected starting values in the optimization algorithm. These initial ranges can be guided by fitting a (time-independent) shared frailty Cox model to obtain suitable bounds (see Section \ref{Application} and Appendix \ref{app:example_time_unvarying} for further details).

Each of the five parameter types in $\boldsymbol{p}$ form a so-called \textit{category}, with the lower bounds for each category collected in the vector
$category_{min} = [min_{\boldsymbol{\phi}}, min_{\boldsymbol{\beta}}, min_{\mu_1}, min_{\nu}, min_{\boldsymbol{\gamma}}]$,
and the upper bounds within the vector 
$category_{max} = [max_{\boldsymbol{\phi}}, max_{\boldsymbol{\beta}}, max_{\mu_1}, max_{\nu}, max_{\boldsymbol{\gamma}}]$.

\subsection{Computation of the model log-likelihood}\label{computation_loglik}

Two functions are employed for the computation of the model log-likelihood.
Specifically, we can observe in Eq. (\ref{ll_AdPaik}) that, exploiting the property of groups independence, the overall log-likelihood function $ll$ can be computed summing up the groups log-likelihood $ll_j$ as $ll = \sum_{j=1}^N ll_j$ with 
\begin{align}
    ll_j &= \sum_{i,k} d_{ijk}( \boldsymbol{\beta}^T \boldsymbol{x}_{ij} +\phi_k) - \frac{\mu_1}{\nu}\text{log}(1+\nu A_{j..}) + \sum_{k} \left[ \frac{-\mu_2}{\gamma_k} \text{log}(1+\gamma_kA_{j.k})\right] + \nonumber\\
    &+ \sum_{k} \left[ \text{log}\left( \sum_{l=0}^{d_{j.k}}\binom{d_{j.k}}{l} \frac{\Gamma(\mu_2/\gamma_k+d_{j.k}-l)}{\Gamma(\mu_2/\gamma_k)} \frac{\Gamma(\mu_1/\nu+l)}{\Gamma(\mu_1/\nu)} \frac{(A_{j.k}+1/\gamma_k)^{(l-d_{j.k})}}{(A_{j..}+1/\nu)^l} \right) \right]. \label{ll_j_AdPaik}
\end{align} 

Through Algorithm~\ref{algorithm:model_ll_j}, the group log-likelihood $ll_j$ in (\ref{ll_j_AdPaik}) is computed, and then employed by Algorithm \ref{algorithm:model_ll} to compute the full model log-likelihood. This second algorithm requires in input
$\boldsymbol{p}$ as in (\ref{parameters_disposition}) (with each category defined in its range), dataset $D$, 
unit group membership $i$-$group$ (i.e., a column vector where each row contains the group a unit belongs to), partitioned time-domain $T_{\text{axis}}$ and matrices $d_{ijk}$ and $e_{ijk}$.

\begin{minipage}{0.9\textwidth}
\begin{algorithm}[H] 
    \caption{Computation of model \textit{group} log-likelihood}
    \label{algorithm:model_ll_j}
    \begin{algorithmic}[1]
    \REQUIRE {$\boldsymbol{p}$, $D_j$, $d_{ijk}$, $e_{ijk}$}
    \STATE Extract from inputs: number of regressors ($R$), parameters ($\sum \boldsymbol{n_p}$), groups ($N$) and intervals ($L$) of $T_{\text{axis}}$.
    \STATE Extract parameters categories from $\boldsymbol{p}$ and save them into $\boldsymbol{\phi}, \boldsymbol{\beta}, \mu_1, \nu, \boldsymbol{\gamma} \to\mu_2 = 1 - \mu_1$. 
    \STATE Extract variables $A_{ijk}, A_{ik}, A_i, d_{ik}$.
    \STATE Initialize $ll_j=0$
    \STATE Compute the terms of $ll_j$ in (\ref{ll_j_AdPaik}) and sum them together into $ll_j$.
    \RETURN {$ll_j$}
    \end{algorithmic}
\end{algorithm} 
\end{minipage}
\begin{minipage}{0.9\textwidth}
\begin{algorithm}[H] 
    \caption{Computation of model log-likelihood}
    \label{algorithm:model_ll}
    \begin{algorithmic}[1]
    \REQUIRE {$\boldsymbol{p}$, $D$, $i$-$group$, $T_{\text{axis}}$, $d_{ijk}$, $e_{ijk}$}
    \STATE Extract from inputs: number of regressors ($R$), parameters ($\sum \boldsymbol{n_p}$), groups ($N$) and intervals ($L$) of $T_{\text{axis}}$.
    \STATE Initialize global log-likelihood $ll$ = 0
    \FOR {$\text{group } j = 1,\dots,N$}
    \STATE{Create the inputs considering only units in group $j$ ($I_j$).}
    \STATE{Compute model log-likelihood $ll_j$ with group-related-variables.}
    \STATE{Sum $ll_j$ to $ll$.}
    \ENDFOR
    \RETURN {$ll$}
    \end{algorithmic}
\end{algorithm} 
\end{minipage}

\subsection{Constrained Maximization of the log-likelihood}\label{max_like}

To maximize the log-likelihood function under parameter constraints, we employ an adaptation of Powell's optimization method for multidimensional spaces \citep{press2007numerical}.
This method takes a set of $\sum \boldsymbol{n_p}$ linearly independent directions and minimizes (maximizes) the function along one direction at a time, using a one-dimensional optimization method (Brent’s method).
The optimization begins at an initial guess for each of the $\sum \boldsymbol{n_p}$ components of $\boldsymbol{p}$, randomly chosen within their five associated pre-defined ranges according to the category each parameter belongs to.
The log-likelihood function is then sequentially optimized along a series of directions, each corresponding to one of the $\sum \boldsymbol{n_p}$ components in $\boldsymbol{p}$.
It proceeds until it completes all the $\sum \boldsymbol{n_p}$ directions and then repeats the same procedure but using another set of ordered directions.

The optimization process continues iteratively until one of two conditions is met: (1) convergence, where the change in the log-likelihood between successive iterations falls below a predefined tolerance $tol_{ll}$, or (2) achievement of the maximum number of iterations $n_{\text{total-run}} = n_{\text{run}} + n_{\text{extra-run}}$, where $n_{\text{run}} = \sum \boldsymbol{n_p}$ and $n_{\text{extra-run}}$ is an input parameter that indicates number of sets of ordered directions according to which the function must be optimized. If convergence is achieved, the process halts early; otherwise, the algorithm continues until all iterations are completed. The status of the algorithm is recorded using the binary variable $Status_{alg}$, which indicates whether convergence was reached (TRUE) or if the maximum number of iterations was exhausted without convergence (FALSE).

We here point out that since the optimization is performed separately for each component $\overline{p}$ of $\boldsymbol{p}$, we need to record the associated index $i_{\overline{p}}$ indicating the position of $\overline{p}$ in $\boldsymbol{p}$ (e.g., if we optimize the log-likelihood function with respect to $\beta_1$, then: $\overline{p} = \beta_1$ and $i_{\overline{p}} = (L+1)$).
This means that when the optimization is performed with respect to a single direction, we also need to specify which is the parameter we are referring to (i.e., $\overline{p}$) and which is its position inside $\boldsymbol{p}$ (i.e., $i_{\overline{p}}$), so that we can identify it correctly. Therefore, we are required to implement another version of both Algorithms \ref{algorithm:model_ll_j} and \ref{algorithm:model_ll}, in a way that the first and the second arguments are the parameter $\overline{p}$ and its associated index $i_{\overline{p}}$.
In detail, both algorithms are modified in the input variables list,  and the algorithm itself is also modified internally, and the assignment $\boldsymbol{p}[i_{\overline{p}}] = \overline{p}$ is executed, so that the current value of that parameter $\overline{p}$ is updated inside. No further changes are necessary and, for this reason, the new implementations are not here reported.

To perform the one-dimensional optimization, we adopt the \textit{Brent's method} \citep{press2007numerical} that uses a combination of \textit{golden section search} and \textit{successive parabolic interpolation} to locate the minimum (maximum) in a precise interval, whose extrema need to be provided. This method is already implemented in \proglang{R} by the functions \texttt{optimize()} and \texttt{optim()} \citep{R}, with the option \lq\lq Brent".
In this way, the global multidimensional optimization method becomes constrained.

Algorithm \ref{algorithm:optimization_phase} reports the pseudo-code for the constrained optimization of the log-likelihood function in multiple dimensions. It requires several inputs, each of them addressed in Section \ref{package}, and it outputs optimized parameters, posterior frailty estimates, and measures of model accuracy that will be addressed in Subsection \ref{outputs}.


\begin{minipage}{0.9\textwidth}
\renewcommand*\footnoterule{%
\kern -6pt}
\begin{algorithm}[H]
    \caption{Constrained optimization in multidimension}
    \label{algorithm:optimization_phase}
    \begin{algorithmic}[1]
    \REQUIRE {$formula$, $data$, $T_{\text{axis}}$, $category_{min}, category_{max}$, $full_{sd}$, 
    $n_{\text{extra-run}}$, $tol_{ll}$, $tol_{optim}$, $h_p$}
    \STATE Extract $D$, $i$-$group$, $t_{ij}$ from $data$ using $formula$.
    \STATE Compute $\sum \boldsymbol{n_p}$, extended vector $params_{min}, params_{max}$
    from $category_{min}, category_{max}$. \\Build random $\boldsymbol{p}$, with values in [$params_{min}, params_{max}$].
    \STATE Build $d_{ijk}$ and $e_{ijk}$.
    \STATE Build $n_{\text{total-run}}$ sets $X_{total-run,dir}$ of possible directions.
    \STATE Set $run$ = 1, $actual_{tol_{ll}}$ = 1, $ll_{optimal} = -1e100$, $Status_{alg} = TRUE$.
    \WHILE {($run <= n_{\text{total-run}}$ and $actual_{tol_{ll}} > tol_{ll}$)}
    \FOR{direction $\boldsymbol{x}_{\overline{p}} \in X_{run,dir}$}
    \STATE {Fix $\boldsymbol{x}_{\overline{p}}$, associated parameter $\overline{p}$} and index $i_{\overline{p}}$. 
    \STATE {Maximize log-likelihood $ll$ (computed through $2^{\text{nd}}$ version of Algorithm \ref{algorithm:model_ll})  \\along $\boldsymbol{x}_{\overline{p}}$ wrt $\overline{p}$, through \textsl{optimize}.} 
    \STATE {Save uni-dimensional maximum $\hat{\overline{p}}$ in $\boldsymbol{p}[i_{\overline{p}}] \to \hat{\boldsymbol{p}}$.} 
    \ENDFOR
    \STATE Store $\hat{\boldsymbol{p}}$ and $ll_{current}(\hat{\boldsymbol{p}})$.
    \STATE Update $actual_{tol_{ll}} = |ll_{optimal} - ll_{current}|.$ 
    \IF{$ll_{optimal} < ll_{current}$}
        \STATE{$ll_{optimal} = ll_{current}$}
    \ENDIF   
    \STATE $run += 1$. 
    \ENDWHILE
    \IF{$run == n_{\text{total-run}}$}
        \STATE{$Status_{alg} = FALSE$}
    \ENDIF
    \STATE Take $\hat{\boldsymbol{p}}$ associated to $ll_{optimal}$ $\to \hat{\boldsymbol{p}}_{optimal}$. \\
    Compute $\boldsymbol{se}_{optimal}$ and 
    $CI(\hat{\boldsymbol{p}}_{optimal})$ employing Algorithm~\ref{algorithm:se_parameters} and (\ref{ci_params}), respectively.
    \STATE {Compute frailty standard deviation $sd$, considering value of $full_{sd}$}.
    \STATE {Compute posterior frailty estimates $\hat{Z}_{jk}$ and $CI(\hat{Z}_{jk})$ $\forall j,k$, 
    employing (\ref{post_frailty_estimates}), (\ref{ci_post_frailty_estimates}) and Algorithm~\ref{algorithm:post_frailty_estimates}}.
    \STATE Compute $AIC$ as specified in (\ref{eq:AIC}).    
    \RETURN $\sum \boldsymbol{n_p}$, $N$, $ll_{optimal}$, $AIC$, $Status_{alg}$, $\hat{\boldsymbol{p}}_{optimal}$, $\boldsymbol{se}_{optimal}$, \\
    $CI(\hat{\boldsymbol{p}}_{optimal})$, $sd$, $\hat{\boldsymbol{Z}}$ and $CI(\hat{\boldsymbol{Z}})$.
    \end{algorithmic}
\end{algorithm}
\end{minipage}
%
%

\subsubsection{1D analysis of the log-likelihood}
\label{sec:1danalysis_methods}

To analyze the behaviour of the log-likelihood function with respect to a single $\overline{p}$ in $\mathbf{p}$, we define a support function 
that performs one-dimensional optimization. 
Specifically, this function has a two-fold use: on one side, it can be used to identify the existence range of one model parameter or to shrink a previously identified one, providing information to be employed in the multi-dimensional optimization phase; on the other side, it can be used to study the behaviour of the log-likelihood function with respect to one parameter, allowing to verify whether the estimated parameter indeed maximizes the function. \\\\
%
In this function, the log-likelihood is maximized with respect to $\overline{p}$ while keeping the other parameters either fixed to default/optimal values or allowed to be randomly assigned within their respective ranges.

The approach involves generating $n_{points}$ random values $\overline{p}_1, \overline{p}_2, \dots, \overline{p}_{n{points}}$ within the range of $\overline{p}$, then computing the log-likelihood values $ll_i$ relative to $\overline{p}_i$, for each $i=1,\ldots,n_{points}$. The points $(\overline{p}_i, ll_i)$ are plotted along with the optimized pair $(\overline{p}_{optimal}, ll_{optimal})$ to visualize the trend of the one-dimensional log-likelihood function.
Since we perform a one-dimensional optimization of the log-likelihood function, we specify the parameter $\overline{p}$ and its index $i_{\overline{p}}$, applying the second version of both Algorithms~\ref{algorithm:model_ll_j} and \ref{algorithm:model_ll}.
The pseudo-code is reported in Algorithm \ref{algorithm:1D_optimization_ll}.
Given in input the parameter $\overline{p}$, it returns its optimized value along with its corresponding log-likelihood. \\
\begin{minipage}{0.9\textwidth}
\renewcommand*\footnoterule{}
\begin{algorithm}[H] 
    \caption{1D optimization of log-likelihood function}
    \label{algorithm:1D_optimization_ll}
    \begin{algorithmic}[1]
    \REQUIRE {$formula$, $data$, $T_{\text{axis}}$, $\overline{p}$, $category_{min}$, $category_{max}$, $flag_{\hat{\boldsymbol{p}}}$, $\boldsymbol{p}$, $n_{rep}$, $tol_{optimize}$, $flag_{plot}$, $n_{points}$}
    \STATE Extract $D$, $i$-$group$, $t_{ij}$ from $data$ using $formula$.
    \STATE Compute $\sum \boldsymbol{n_p}$ and define $\boldsymbol{p} = \boldsymbol{0}$.
    \STATE Compute extended vector $params_{min}, params_{max}$ from $category_{min}, category_{max}$.
    \STATE Build $d_{ijk}$ and $e_{ijk}$.
    \FOR{$j = 1, \dots, n_{rep}$}
    \IF{$flag_{\hat{\boldsymbol{p}}} == 1$} 
        \STATE $\boldsymbol{p} = \hat{\boldsymbol{p}}$ and $\boldsymbol{p}[\overline{pp}] = unif(1, params_{min}[\overline{pp}], params_{max}[\overline{pp}])$\\
    \ELSE
        \STATE $\boldsymbol{p} = unif(n_p, params_{min}, params_{max})$
    \ENDIF
    \STATE Optimize $ll$ with respect to $\overline{p}$. \\Save results into ($\overline{p}_{optimal}$, $ll_{optimal}$).
    \IF{$flag_{plot} == 1$}
        \STATE Generate $n_{points}$ $\overline{p}_i$, save $\boldsymbol{p}[\overline{pp}] = \overline{p}_i$
        and evaluate $ll_i = ll(\boldsymbol{p})$. \\
        Plot $(\overline{p}_i, ll_i)$ $\forall i$.
    \ENDIF
    \ENDFOR
    \RETURN ($\overline{p}_{optimal}$, $ll_{optimal}$) $\forall j$.
    \end{algorithmic}
\end{algorithm} 
\end{minipage}

\subsection{Outputs computation}\label{outputs}

We here consider $\hat{\boldsymbol{p}}_{optimal}$ and $ll_{optimal}$ given in output by the log-likelihood maximization algorithm described above, and report the computation of the other non-trivial outputs in Algorithm \ref{algorithm:optimization_phase}, i.e., AIC; parameters standard errors $\boldsymbol{se}_{optimal}$ and their confidence interval $CI(\hat{\boldsymbol{p}}_{optimal})$; frailty standard deviation $sd$, posterior fraity estimates $\hat{\boldsymbol{Z}}$ and $CI(\hat{\boldsymbol{Z}})$.
 
\subsubsection{Model goodness of fit}
To evaluate the goodness of fit of the model we employ the \textit{Akaike information criterion} (AIC) \citep{akaike1998information, bozdogan1987model}. 
Given the value of the optimized log-likelihood $ll_{optimal}$ and the number of unknown parameters $\sum \boldsymbol{n_p}$ characterizing the model, the AIC can be computed as: 
\begin{equation}\label{eq:AIC}
    AIC = 2 \sum \boldsymbol{n_p} - 2 ll_{optimal}
\end{equation}
This index quantifies the loss of information of the model under consideration and it can be used to compare the performance of different models: the lower the AIC, the better the associated model.
\subsubsection{Parameters standard error}
To quantify the accuracy of our estimates, we compute the \textit{standard error} of every parameter $\hat{\boldsymbol{p}}_{optimal}$ as:
\begin{equation*}
    \boldsymbol{se}_{i_p}(\hat{\boldsymbol{p}}_{optimal}) = 1 / \sqrt{I_{i_p}(\hat{\boldsymbol{p}}_{optimal})} 
\end{equation*}
where $I_{i_p}(\hat{\boldsymbol{p}}_{optimal})$ is the $p$-th element of the diagonal of the \textit{information matrix}, computed as the opposite of the \textit{Hessian matrix}, evaluated at the $\hat{\boldsymbol{p}}_{optimal}$, as $I(\hat{\boldsymbol{p}}_{optimal}) = - H(\hat{\boldsymbol{p}}_{optimal})$.\\

While the \texttt{optim()} function in \proglang{R} can compute the Hessian during each iteration of the log-likelihood maximization, doing so repeatedly for each parameter across all runs (i.e., $(n_{\text{total-run}} \cdot \sum \mathbf{n_p})$ times) would be computationally expensive and inefficient in terms of both time and memory. 
To address this, we employ an auxiliary algorithm that approximates only the diagonal of the Hessian matrix at the end of the optimization phase. This approach leverages a \textit{centered finite difference scheme} for second-order accuracy, which is more efficient since the standard error formula requires only the diagonal elements. This method is both memory-efficient and faster than calculating the full Hessian at every iteration.
Algorithm~\ref{algorithm:se_parameters} summarizes the procedure.  The final optimized parameters and a small discretization step $h_p$ are required in input.

\begin{minipage}{0.9\textwidth}
\renewcommand*\footnoterule{%
\kern -6pt}
\begin{algorithm}[H] 
    \caption{Computation of parameters standard error}
    \label{algorithm:se_parameters}
    \begin{algorithmic}[1]
    \REQUIRE {$\hat{\boldsymbol{p}}_{optimal}$, $D$, $i$-$group$, $T_{\text{axis}}$, $d_{ijk}$, $e_{ijk}$, $h_p$.}
    \STATE Compute $\sum \boldsymbol{n_p}$ and initialize $\boldsymbol{se}_{optimal} = \boldsymbol{0}$.
    \FOR {$pp = 1, \dots, n_p$}
    \STATE{Extract $\hat{p} = \hat{\boldsymbol{p}}[pp]$ and define $\hat{p}_{+} = (\hat{p} + h_p)$, $\hat{p}_{-} = (\hat{p} - h_p)$}.
    \STATE{Re-assign $\hat{\boldsymbol{p}}_{+}[pp] = \hat{p}_{+}$, $\hat{\boldsymbol{p}}_{-}[pp] = \hat{p}_{-}$}
    \STATE Compute (employing Algorithm \ref{algorithm:model_ll}) $ll_{+} = ll(\hat{\boldsymbol{p}}_{+})$, $ll_{-} = ll(\hat{\boldsymbol{p}}_{-})$ and $ll = ll(\hat{\boldsymbol{p}})$.
    \STATE Compute $H(\hat{p}) = (ll_{-} - 2ll + ll_{+})/h_p^2$
    \STATE Compute $\boldsymbol{se}_{optimal}(\hat{p}) = 1/\sqrt{-H(\hat{p})}$
    \ENDFOR
    \RETURN {$\boldsymbol{se}_{optimal}$}
    \end{algorithmic}
\end{algorithm} 
\end{minipage}
\subsubsection{Posterior frailty estimates, variances and confidence interval}
Once the optimization phase is executed and the optimal parameters recovered, we can derive a posteriori the frailty estimates, their variances and confidence interval, as discussed in Section~\ref{adapted_paik_eam}, but we first need to introduce $N_j(I_k)$, $\hat{H}_{j,\bullet}(I_k)$ and $\hat{H}_{j,\bullet}$. \\First of all, $N_j(I_k)$ quantifies the number of events happened in each group $j$ and interval $I_k$. It can be evaluated starting from the time-to-event $t_{ij}$ and controlling whether it is less than the end of the follow-up and, if so, in which interval $I_k$ it falls. Indeed, thanks to the modification of the variable $t_{ij}$, we know that a time-to-event in the follow-up implies an event.\\
Concerning the cumulative hazard function, the unit at-risk indicator $Y_{ij}$ is a vector of length equal to the number of intervals and each element $Y_{ij}(I_k)$ is a binary variable indicating if unit $i$ in group $j$ is at risk ($Y_{ij}(I_k) = 1$) of facing the event in interval $I_k$. In detail:
\begin{equation*}
    Y_{ij}(I_k) =\begin{cases}  1 & \forall k < \overline{k}, \text{ such that } t_{ij} \notin I_{\overline{k}} \\ 
    0 &  \forall k \geq \overline{k}, \text{ such that } t_{ij} \in I_{\overline{k}}
\end{cases}
\end{equation*}
Then, the unit cumulative hazard function $\hat{H}_{ij}(I_k)$ can be computed as: 
\begin{equation*}
    \hat{H}_{ij}(I_k) = e_{ijk} Y_{ij}(I_k) exp(\hat{\boldsymbol{\beta}}^T \boldsymbol{x}_{ij}  + \hat{\phi}_k)
\end{equation*}
where both the regressors $\hat{\beta}_r$ $\forall r$ and the interval baseline log-hazard $\hat{\phi}_k$ $\forall k$ are the optimal ones.\\
Finally, the contribute of all units in the same group and interval is summed to obtain $\hat{H}_{j,\bullet}(I_k)$, as $\hat{H}_{j,\bullet}(I_k) = \sum_{i \in I_j} \hat{H}_{ij}(I_k)$. In the case of time-independent estimate of $\hat{\alpha}_j$: $\hat{H}_{j,\bullet} = \sum_k \hat{H}_{j,\bullet}(I_k)$ and $N_j = \sum_k N_j(I_k)$.\\\\
In the end, we substitute the just mentioned quantities and the optimal parameters inside (\ref{post_frailty_estimates}), (\ref{post_frailty_variance}) and (\ref{ci_post_frailty_estimates}), as illustrated in Algorithm~\ref{algorithm:post_frailty_estimates}, to compute both the posterior frailty estimates, the posterior frailty variance and the posterior frailty confidence interval.

\begin{minipage}{0.9\textwidth}
\renewcommand*\footnoterule{%
\kern -6pt}
\begin{algorithm}[H] 
    \caption{Computation of posterior frailty estimates and posterior frailty variance}
    \label{algorithm:post_frailty_estimates}
    \begin{algorithmic}[1]
    \REQUIRE {$\hat{\boldsymbol{p}}_{optimal}$, $D$, $t_{ij}$, $i$-$group$, $T_{\text{axis}}$}
    \STATE Extract $\{\phi_1, \dots, \phi_L, \beta_1, \dots, \beta_R, \mu_1, \nu, \gamma_1, \dots, \gamma_L, \mu_2 = 1 - \mu_1\}$ from $\hat{\boldsymbol{p}}_{optimal}$ \\
    Compute $par_{1, \alpha} = \mu_1 / \nu$, $par_{2, \alpha} = 1/\nu$, $par_{1,k, \epsilon} = \mu_2 / \gamma_k$, $par_{2,k, \epsilon} = 1/ \gamma_k$.
    \FOR {$j = 1, \dots, N$}
    \STATE{Extract units $i$ in cluster $j$ ($I_j$)}.
    \FOR {$k = 1, \dots, L$} 
    \STATE{Compute $N_{jk} = N_j(I_k)$}.
    \FOR {$i \in I_j$}
    \STATE{Compute $e_{ijk}$, $Y_{ijk}$, $H_{ijk} = e_{ijk} Y_{ijk} \text{exp}(\hat{\phi}_k + \boldsymbol{x}_{ij} \hat{\boldsymbol{\beta}})$}
    \ENDFOR
    \STATE{Compute $H_{jk} = \sum_{i \in I_j} H_{ijk}$, $\hat{\epsilon}_{jk} = (par_{1,k, \epsilon} + N_{jk}) / (par_{2,k, \epsilon}+ H_{jk})$} and \\$Var(\hat{\epsilon}_{jk}) = \hat{\epsilon}_{jk} / (par_{2,k, \epsilon}+ H_{jk})$
    \ENDFOR
    \STATE{Compute $N_j = \sum_k N_{jk}$, $H_j = \sum_k H_{jk}$, $\hat{\alpha}_j = (par_{1,\alpha} + N_j)/(par_{2,\alpha} + H_j)$} and \\$Var(\hat{\alpha}_j) = \hat{\alpha}_j / (par_{2,\alpha} + H_j).$
    \ENDFOR
    \STATE{Determine $max_{j}{\alpha_j}$ and $max_{j,k}{\epsilon_{jk}}$}.\\
    Compute $\hat{Z}_{jk}$ as in (\ref{post_frailty_estimates}), $var(\hat{Z}_{jk})$ as in (\ref{post_frailty_variance}) and  $CI(\hat{Z}_{jk})$ as in (\ref{ci_post_frailty_estimates}).
    \RETURN {$\hat{Z}_{jk}, var(\hat{Z}_{jk})$ $\forall j,k$}
    \end{algorithmic}
\end{algorithm} 
\end{minipage}

\section{Syntax and Implementation details}\label{package} 

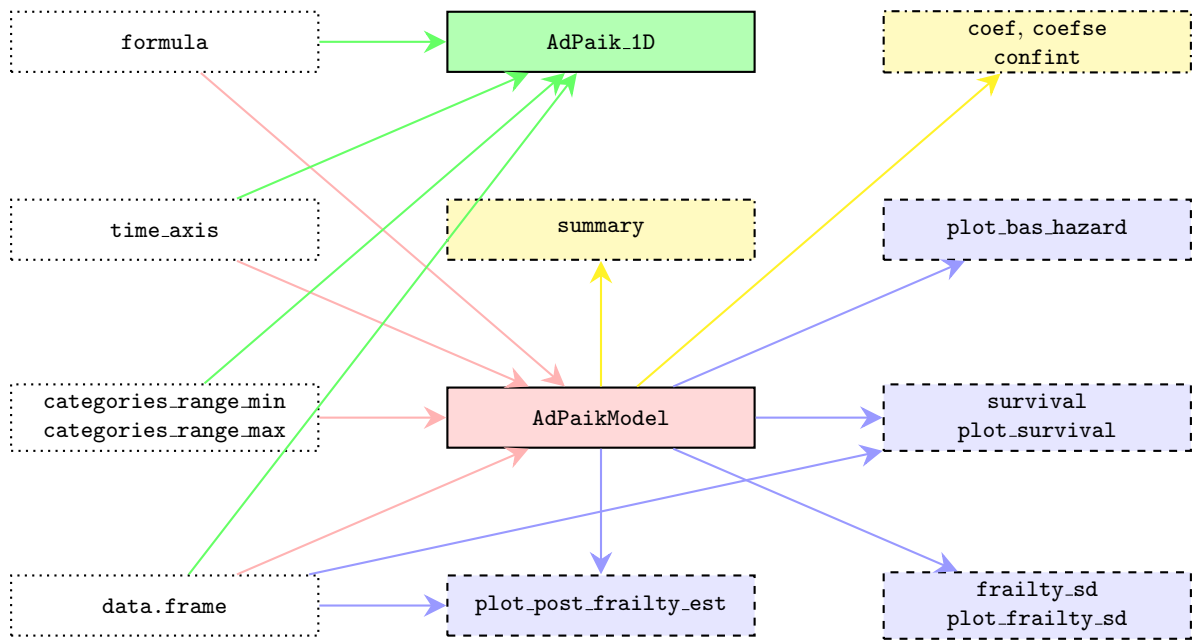
\begin{figure}[]
    \centering
\begin{tikzpicture}[node distance=2.8cm, auto, thick]

    \tikzstyle{dataframe} = [rectangle, draw=black, fill=white, text centered, minimum height=0.8cm, minimum width=2cm, dotted, text width=3.85cm, font=\small]
    \tikzstyle{model} = [rectangle, draw=black, fill=red!15, text centered, minimum height=0.8cm, minimum width=2cm, text width=3.85cm, font=\small]
    \tikzstyle{summary} = [rectangle, draw=black, fill=yellow!30, text centered, minimum height=0.8cm, minimum width=2cm, dashdotted, text width=3.85cm, font=\small]
    \tikzstyle{plot} = [rectangle, draw=black, fill=blue!10, text centered, minimum height=0.8cm, dashed, minimum width=2cm, 
    text width=3.85cm, font=\small]
    \tikzstyle{1d} = [rectangle, draw=black, fill=green!30, text centered, minimum height=0.8cm, minimum width=2cm, 
    text width=3.85cm, font=\small]


    \node[1d] (1d) {\texttt{AdPaik\_1D}};

    \node[summary, below of=1d, yshift=0.3cm] (summary) {\texttt{summary}};
    
    \node[model, below of=summary, yshift=0.3cm] (adpaik) {\texttt{AdPaikModel}}; 

    \node[plot, below of= adpaik, yshift=0.3cm] (plotpost) {\texttt{plot\_post\_frailty\_est}};

    \node[dataframe, left of = 1d, xshift=-3cm] (formula) {\texttt{formula}};
    
    \node[dataframe, left of = summary, xshift=-3cm] (timeaxis) {\texttt{time\_axis}};

    \node[dataframe, left of = adpaik, xshift=-3cm, yshift=0cm] (categories) {\texttt{categories\_range\_min} \\ \texttt{categories\_range\_max}};

    \node[dataframe, left of = plotpost, xshift=-3cm] (data) {\texttt{data.frame}};

    \node[summary, right of = 1d, xshift=3cm] (extr) {\texttt{coef}, \texttt{coefse} \\ \texttt{confint}};
    

    \node[plot, right of = adpaik, xshift=3cm] (plotsurv) {\texttt{survival} \\ \texttt{plot\_survival}};

    \node[plot, right of = summary, xshift=3cm] (plothaz) {\texttt{plot\_bas\_hazard}};
    
    \node[plot, right of=plotpost, xshift=3cm] (plotsd) {\texttt{frailty\_sd} \\ \texttt{plot\_frailty\_sd}};

    \draw[-{Stealth[length=3mm, width=3mm]}, red!30] (data) -- (adpaik);
    \draw[-{Stealth[length=3mm, width=3mm]}, blue!40] (data) -- (plotsurv);
    \draw[-{Stealth[length=3mm, width=3mm]}, red!30] (formula) -- (adpaik);
    \draw[-{Stealth[length=3mm, width=3mm]}, red!30] (categories) -- (adpaik);
    \draw[-{Stealth[length=3mm, width=3mm]}, red!30] (timeaxis) -- (adpaik);
    \draw[-{Stealth[length=3mm, width=3mm]}, blue!40] (data) -- (plotpost);
    \draw[-{Stealth[length=3mm, width=3mm]}, blue!40] (adpaik) -- (plotpost);
    \draw[-{Stealth[length=3mm, width=3mm]}, yellow!90] (adpaik) -- (summary);
    \draw[-{Stealth[length=3mm, width=3mm]}, yellow!90] (adpaik) -- (extr);
    \draw[-{Stealth[length=3mm, width=3mm]}, blue!40] (adpaik) -- (plotsd);
    \draw[-{Stealth[length=3mm, width=3mm]}, blue!40] (adpaik) -- (plothaz);
    \draw[-{Stealth[length=3mm, width=3mm]}, green!60] (formula) -- (1d);
    \draw[-{Stealth[length=3mm, width=3mm]}, green!60] (categories) -- (1d);
    \draw[-{Stealth[length=3mm, width=3mm]}, green!60] (timeaxis) -- (1d);
    \draw[-{Stealth[length=3mm, width=3mm]}, green!60] (data) -- (1d);
    \draw[-{Stealth[length=3mm, width=3mm]}, blue!40] (adpaik) -- (plotsurv);
\end{tikzpicture}
\caption{Diagram illustrating the main functions of the \pkg{TimeDepFrail} package and their interactions. On the left, the nodes with dotted borders represent inputs needed for the main \texttt{AdPaikModel()} function, which fits the model, represented with continuous borders. The \texttt{AdPaikModel()} function produces an output that can be used as input for the other functions. \texttt{AdPaik\_1D()} is used as a support function for likelihood evaluation.
The function with dashed-dotted borders is \texttt{summary()}, which provides tools for model evaluation, as well as the other extractors \texttt{coef(), coefse(), confint()}.
The functions with dashed borders are used to produce estimated objects and their plot: \texttt{plot\_bas\_hazard()}, \texttt{survival()} and \texttt{plot\_survival()}, \texttt{frailty\_sd()} and \texttt{plot\_frailty\_sd()}, and \texttt{plot\_post\_frailty\_est()}.}
    \label{fig:diagram}
\end{figure}

The \pkg{TimeDepFrail} package is built around functions both for modeling and for plotting.
In Figure \ref{fig:diagram}, the relationships between the different tools and functions in the \pkg{TimeDepFrail} package are shown.

In the following three sections, we describe each of them, comprehensively specifying inputs and outputs.
Furthermore, in Section \ref{syntax_1d}, we describe the syntax of \texttt{AdPaik\_1D()}, a support function suitable for the choice of the range of the parameters and analysis of the 1D log-likelihood.

\subsection{Main model and summary: AdPaikModel(), summary() and plot\_bas\_hazard()}
\label{sec:AdPaikModel_fun}

The \texttt{AdPaikModel()} function has the following syntax:
    \begin{CodeChunk}
\begin{CodeInput}
AdPaikModel(
  formula,
  data,
  time_axis,
  categories_range_min,
  categories_range_max,
  flag_fullsd = TRUE,
  n_extrarun = 60,
  tol_ll = 1e-06,
  tol_optimize = 1e-06,
  h_dd = 0.001,
  print_previous_ll_values = c(TRUE, 3),
  level = 0.95,
  verbose = FALSE
)
\end{CodeInput}
\end{CodeChunk}

In Figure \ref{fig:diagram}, on the left column, we indicate the non-default inputs required by the \texttt{AdPaikModel()}. More in detail, 
\begin{itemize}
    \item the \texttt{formula} expresses the relationship between the unit survival times, covariates, and group/cluster membership, expressed as \texttt{$t_{ij} \sim$ covariates + cluster(group)}, where $t_{ij}$ we recall denotes the time-to-event variable and the \texttt{cluster()} term is introduced to distinguish with respect to the variable \texttt{group} under which units are grouped; 
    
    \item \texttt{data} ($D$ in previous section) is a \texttt{data.frame} object where the rows represent the units and the columns the regressors (also other variables that will not be considered during the application may be contained), and where numerical variables must be standardized, while categorical variables must be converted into dummy variables;
    
    \item \texttt{time\_axis} ($T_{\text{axis}}$ in previous section) is the vector representing the partitioned time domain;
    
    \item \texttt{categories\_range\_min} and \texttt{categories\_range\_max} are the five-dimensional vectors containing the categories bounds, required for constrained optimization.
\end{itemize}

Furthermore, among the default inputs, we find:
\begin{itemize}
    \item \texttt{flag\_fullsd} ($full_{sd}$ in previous section), a logical value, if TRUE, the full frailty standard deviation is computed, otherwise the partial one that keeps into account only the time-dependent component. Defaults to TRUE.
    \item \texttt{n\_extrarun} ($n_{\text{extra-run}}$ in previous section), total number of runs (iterations) are get summing to the number of parameters the number of extrarun.
    \item \texttt{tol\_ll} ($tol_{ll}$ in previous section), tolerance on the log-likelihood value.
    \item \texttt{tol\_optimize} ($tol_{optim}$ in previous section), the desired accuracy on the maximum to be found, a value is required internally by both functions \textsl{optimize()} and \textsl{optim()} \citep{R} in \proglang{R};
    \item \texttt{h\_dd}  ($h_p$ in previous section), the value of the discretization step used for the numerical approximation of the second derivative of the log-likelihood (see next subsection \ref{outputs})
    \item \texttt{print\_previous\_ll\_values} allows to specify if we want to print the previous values of the log-likelihood function. This can be useful for controlling that the optimization procedure is proceeding in a monotone way and it does not oscillate. This argument is composed of two elements: TRUE/FALSE if we want or not to print the previous values and how many values we want to print on the console. Default is (TRUE, 3), so that only the previous 3 values of the log-likelihood are printed.
    \item \texttt{level}, a numeric value representing the confidence level for the optimal parameters (default is 0.95 for 95\% confidence).
    \item \texttt{verbose}, a logical value, if TRUE, detailed progress messages will be printed to the console. Defaults to FALSE.
\end{itemize}

At the end of the optimization phase, several quantities are estimated and the function returns an S3 object of class \lq AdPaik', described in the \pkg{TimeDepFrail} package documentation of \texttt{AdPaikModel()}, composed of several elements:


\begin{itemize}
\item \texttt{formula}: formula object provided in input by the user.
\item \texttt{Regressors}: categorical vector of length R, with the name of the regressors.
They could be different from the original covariates of the dataset in case of categorical covariates.
Indeed, each categorical covariate with $n$ levels needs to be transformed into ($n-1$) dummy variables and, therefore, ($n-1$) new regressors.
\item \texttt{NRegressors}: number of regressors $R$.
\item \texttt{ClusterVariable}: name of the variable with respect to which the individuals can be grouped.
\item \texttt{NClusters}: number of clusters/groups N.
\item \texttt{TimeDomain}: partitioned time domain $T_{\text{axis}}$.
\item \texttt{NIntervals}: number of intervals of the time-domain L.
\item \texttt{NParameters}: number of parameters of the model $n_p = 2L + R + 2$.
\item \texttt{ParametersCategories}: Numerical vector of length 5, containing the numerosity of each parameter category.
\item \texttt{ParametersRange}: S3 object of class \lq ParametersRange\rq, containing \texttt{ParametersRangeMin} and  \texttt{ParametersRangeMax}, two numerical vectors of length $n_p$, giving the minimum and the maximum range of each parameter, respectively.
\item \texttt{Loglikelihood}: value of the maximized log-likelihood function, at the optimal estimated parameters.
\item \texttt{AIC}: \lq Akaike Information Criterion\rq: it can be computed as $AIC = 2n_p - 2ll_{\text{optimal}}$.
It quantifies the loss of information related to the model fitting and output.
The smaller, the less the loss of information and the better the model accuracy.
\item \texttt{Status}: Logical value. TRUE if the model reaches convergence, FALSE otherwise.
\item \texttt{NRun}: Number of runs necessary to reach convergence. If the model does not reach convergence, such number is equal to the maximum number of imposed runs.
\item \texttt{OptimalParameters}: numerical vector of length $n_p$, containing the optimal estimated parameters (the parameters that maximize the log-likelihood function).
\item \texttt{StandardErrorParameters}: numerical vector of length $n_p$, corresponding to the standard error of each estimated parameters.
\item \texttt{ParametersCI}: S3 object of class \lq ParametersCI\rq, composed of two numerical vector of length equal to $n_p$: the left and right confidence interval of each estimated parameter of given \texttt{level}.
\item \texttt{BaselineHazard}: numerical vector of length equal to $L$, containing the baseline hazard step-function.
\item \texttt{FrailtyDispersion}:  S3 object of class \lq FrailtyDispersion\rq, containing two numerical vectors of length equal to $L$ with the standard deviation and the variance of the frailty.
\item \texttt{PosteriorFrailtyEstimates}: S3 object of class \lq PFE.AdPaik\rq. See details.
\item \texttt{PosteriorFrailtyVariance}: S3 object of class \lq PFV.AdPaik\rq. See details.
\item \texttt{PosteriorFrailtyCI}: S3 object of class \lq PFCI.AdPaik\rq. See details.
\end{itemize}

It is also possible to employ \texttt{summary()}, 
which requires as input the S3 object output of \texttt{AdPaikModel()} (that for convenience, from now on, we name \texttt{result}), as follows \texttt{summary(result)}, and summarizes the most important information related to the dataset (number of units, number of regressors, number of intervals, number of clusters), the model call (number of parameters) and the model output (optimal log-likelihood value and AIC). An example is given in Section \ref{Application}.
Furthermore, to better extract categories related to $\boldsymbol{\phi}$, $\boldsymbol{\beta}$, $\mu$, $\nu$, $\boldsymbol{\gamma}$ from \texttt{OptimalParameters}, \texttt{StandardErrorParameters} and \texttt{ParametersCI}, we provided three extractors \texttt{coef()}, \texttt{coefse()} and \texttt{confint()}, respectively, following the logic in Table~\ref{table_extraction_param}.

\begin{table}\small
\centering 
\begin{adjustbox}{width=1\textwidth}
\begin{tabular}{@{}ccccccc@{}}
\toprule
\textbf{Vector} & \textbf{Param Category} & \textbf{Extended Parameters} & \textbf{Numerosity} & \textbf{Extraction from Model Output} \\ 
\midrule
\multirow{5}{*}{$\boldsymbol{p}$} 
& $\boldsymbol{\phi}$       & $[\phi_1, \dots, \phi_L]$        & $L$  & \texttt{\$OptimalParameters[1:L]} \\
& $\boldsymbol{\beta}$      & $[\beta_1, \dots, \beta_R]$      & $R$  & \texttt{\$OptimalParameters[(L+1):(L+R)]} \\
& $\mu$                     & {}                               & $1$  & \texttt{\$OptimalParameters[L+R+1]} \\
& $\nu$                     & {}                               & $1$  & \texttt{\$OptimalParameters[L+R+2]} \\
& $\boldsymbol{\gamma}$     & $[\gamma_1, \dots, \gamma_L]$    & $L$  & \texttt{\$OptimalParameters[(L+R+3):(2*L+R)]} \\
\bottomrule
\end{tabular}
\end{adjustbox}
\caption{Summary of parameter categories and their extraction from model output.}
\label{table_extraction_param}
\end{table}

Lastly, to visualize the piecewise linear estimated baseline hazard, available in the field \texttt{BaselineHazard},
we implemented a method that takes \texttt{result} (the S3 object of class \lq \texttt{AdPaik}') as input, together with other fields controlling the plot's output details.

\begin{CodeChunk}
\begin{CodeInput}
plot_bas_hazard(
  result,
  xlim = c(min(result$TimeDomain), max(result$TimeDomain)),
  ylim = c(0, max(result$BaselineHazard)),
  xlab = "Time",
  ylab = "Baseline hazard",
  main_title = "Baseline hazard step-function",
  color = "black",
  pch = 21,
  bg = "black",
  cex_points = 0.7
)
\end{CodeInput}
\end{CodeChunk}

\subsection{Frailty estimation: frailty\_sd(), plot\_frailty\_sd() and plot\_post\_frailty\_est()}
\label{sec:functions_frailtysd}

As discussed in Section~\ref{adapted_paik_eam}, the frailty variance includes both time-dependent and constant components of heterogeneity. The standard deviation of each term can be assessed employing 

\begin{CodeChunk}
\begin{CodeInput}
frailty_sd(result, flag_fullsd = TRUE)
\end{CodeInput}
\end{CodeChunk}

where \texttt{flag\_fullsd} is a logical value, if \lq\lq TRUE\rq\rq \, (default) the full variance/standard deviations are computed (i.e., including both the time-dependent and constant spreads).
If set to TRUE, it gives the same result as 
\begin{CodeChunk}
\begin{CodeInput}
R> result$FrailtyDispersion
\end{CodeInput}
\end{CodeChunk}

To plot specific components in each time interval (represented by its mid point), we employ \texttt{plot\_frailty\_sd()}
\begin{CodeChunk}
\begin{CodeInput}
plot_frailty_sd(
  result,
  flag_variance = FALSE,
  flag_sd_external = FALSE,
  frailty_sd = NULL,
  xlim = c(min(result$TimeDomain), max(result$TimeDomain)),
  ylim = NULL,
  xlab = "Time",
  ylab = "Values",
  main_title = NULL,
  pch = 21,
  color_bg = "blue",
  cex_points = 0.7
)
\end{CodeInput}
\end{CodeChunk}

where \texttt{flag\_variance} is a binary flag, if \lq\lq TRUE\rq\rq \, the variance instead of the standard deviation is plotted, \texttt{flag\_sd\_external} is a binary flag, if \lq\lq TRUE\rq\rq \, the standard deviation is passed through \texttt{frailty\_sd}, and \texttt{frailty\_sd} specifies the component of the variance/standard deviation to plot (if none is given, the full one is employed), and the other inputs control the plot's output details.
An example of practical use of the function is given in Section \ref{Application}.


Another useful implemented method is \texttt{plot\_post\_frailty\_est()}, 

\begin{CodeChunk}
\begin{CodeInput}
plot_post_frailty_est(
  result,
  data,
  flag_eps = FALSE,
  flag_alpha = FALSE,
  xlim = NULL,
  ylim = NULL,
  xlab = "Time",
  ylab = "Values",
  main_title = "Posterior frailty estimates",
  cex = 0.7,
  pch_type = rep(21, length(centre_codes)),
  color_bg = rep("black", length(centre_codes)),
  cex_legend = 0.7,
  pos_legend = "topright"
)
\end{CodeInput}
\end{CodeChunk}

where \texttt{data} should be a \texttt{data.frame} in which all variables of the formula object must be found and contained, \texttt{flag\_eps} is a binary flag, if \lq\lq TRUE\rq\rq \,  only the time-dependent posterior frailty estimates are plotted, and \texttt{flag\_alpha} is another binary flag, if \lq\lq TRUE\rq\rq \, only the time-independent posterior frailty estimates are plotted. Note that, it is not possible to have both previous flags \lq\lq TRUE\rq\rq: either one of the two must be true. However, both can be \lq\lq FALSE\rq\rq, and in this case, only the total one is plotted.
Also in this case, we plot a value for each time interval (represented by its mid point).

\subsection{Prediction of survival curves: survival() and plot\_survival()}
\label{sec:survival}

It is also possible to create a dataset where each row corresponds to an individual unit in the dataset, and the columns represent the survival function values over time intervals, with the first column indicating the cluster to which the individual belongs. This can be achieved using the following function:

\begin{CodeChunk}
\begin{CodeInput}
survival(result, data)
\end{CodeInput}
\end{CodeChunk}

which takes in input \texttt{result}, the S3 object of class 'AdPaik' containing model results, and a dataframe \texttt{data} containing covariates used in the model.

This result can then be plotted through the function

\begin{CodeChunk}
\begin{CodeInput}
plot_survival(
  result,
  survival_df,
  lwd = 1,
  xlim = c(min(result$TimeDomain), max(result$TimeDomain)),
  ylim = c(0, 1),
  xlab = "Time",
  ylab = "Values",
  main = "Survival",
  cex = 0.2,
  cexlegend = 0.8
)
\end{CodeInput}
\end{CodeChunk}

where in this case \texttt{survival\_df} is the output of \texttt{survival()}.
Also in this case, we plot a value for each time interval (represented by its mid point).

\subsection{Auxiliary function: AdPaik\_1D()}
\label{syntax_1d}

Lastly, this function is suitable for studying the log-likelihood function from the point of view of a single parameter and, therefore, in a single direction. It performs both the optimization of the log-likelihood with respect to this parameter and the evaluation of the log-likelihood in a specific range of the parameter, while the other parameters can assume a constant predefined value or can be randomly assigned in their existence range. 
Examples of the different usages are presented in Section \ref{1D_analysis}. The syntax is the following: 

\begin{CodeChunk}
    \begin{CodeInput}
AdPaik_1D(
  formula,
  data,
  time_axis,
  index_param_to_vary,
  flag_optimal_params = FALSE,
  optimal_params = NULL,
  categories_range_min,
  categories_range_max,
  n_iter = 5,
  tol_optimize = 1e-06,
  flag_plot = FALSE,
  n_points = 150,
  cex = 0.7,
  cex_max = 0.8,
  color_bg = "black",
  color_max_bg = "red",
  pch = 21
)
    \end{CodeInput}
\end{CodeChunk}
Among the inputs not yet mentioned, we find
\begin{itemize}
    \item \texttt{index\_param\_to\_vary}, the index of the parameter with respect to which the log-likelihood function is studied and analysed (see Table \ref{table_extraction_param}).
    \item \texttt{flag\_optimal\_params} ($flag_{\hat{\boldsymbol{p}}}$ in previous section): if the aim is to study the 1D log-likelihood function keeping fixed the other parameters to their optimal (optimized) value, then this input binary variable should be equal to \lq\lq TRUE\rq\rq.  Otherwise, when studying the log-likelihood with the parameters randomly selected in their ranges, the flag must be equal to \lq\lq TRUE\rq\rq.
    \item \texttt{optimal\_params}: vector of optimal parameters, determined through an entire multi-dimensional maximization of the log-likelihood function. The default value set to \textit{NULL} indicates that no vector is provided and the parameters are randomly extracted in their ranges. 
    \item \texttt{n\_iter} ($n_{rep}$ in previous section): number of times the one-dimensional analysis with respect to the indicated parameter must be executed. 
    If \texttt{flag\_optimal\_params} = \lq\lq TRUE\rq\rq, it is not necessary to perform it several times because the same result will be obtained. Thus, it is more useful in case of randomly generated parameters to observe the trend of the log-likelihood function and evaluate its maximum\footnote{At each iteration, new random parameters are extracted, leading to different maximum points of the log-likelihood function that, combined together, identify a possible existence range for the analysed parameter.}.
    \item \texttt{tol\_{optimize}} ($tol_{optimize}$): tolerance internally used by \textit{optimze} to maximize the log-likelihood function.
    \item \texttt{flag\_plot} ($flag_{plot}$): flag for plotting the log-likelihood trend. In order to observe the trend of the log-likelihood function, this input should be set to \lq\lq TRUE\rq\rq; otherwise, to \lq\lq FALSE\rq\rq.
    \item \texttt{n\_points} ($n_{points}$): in case of \texttt{flag\_plot} = \lq\lq TRUE\rq\rq, this input indicates the number of internal points in which the log-likelihood function must be evaluated, to plot it.
\end{itemize}
%
%


\section{Worked example}\label{Application}

In this section, we present a reproducible example employing data extracted from an administrative database of an Italian university.
In Section \ref{inputs} we consider the non-default inputs for the \texttt{AdPaikModel} function, giving some practical insights in how to proceed.
In Section \ref{model_execution_output}, we detail the model execution and computed parameters. In Section \ref{sec:analysis_of_frailty}, we detail the frailty standard deviation and the posterior frailty estimates. In Section \ref{sec:condsurvfun}, we detail the conditional survival function.
Finally, in Section \ref{1D_analysis}, we describe the use of \texttt{AdPaik\_1D}, a support function suitable for the one-dimensional analysis of the log-likelihood function.

\subsection{Non-default inputs}\label{inputs}

The \texttt{data$\_$dropout} dataset comprises information related to $4448$ students enrolled in 16 degree programs (identified by codes such as $CosA$, $CosB$, $\dots$, $CosP$) in the 2012 academic year. The event of interest is \textit{academic dropout}, or university withdrawal, within bachelor's degree. The \textit{time-to-event} records the semester of dropout with a decimal precision\footnote{Indeed, within this specific university, a student can drop out at any moment of his career.}; if no dropout occurs (censored observation), the time-to-event is set to $6.1$ semesters.

This dataset consists of $4448$ rows (one for each student) and $4$ columns that are \texttt{Gender} (\textit{Male} or \textit{Female}\footnote{Reference level is \textit{Female}, with the regressor \textit{GenderMale}.}), \texttt{CFUP} (number of credits\footnote{\textit{Credito Formativo Universitario (CFU).}} earned by the student by the end of the first semester, max $30$), the \texttt{time-to-event} variable, and degree course membership (\texttt{group}). Higher \texttt{CFUP} is expected to correlate with lower dropout risk.
Below are the first few rows of the dataset:
\begin{CodeChunk}
\begin{CodeInput}
R> data(data_dropout)
R> head(data_dropout)
\end{CodeInput}
\begin{CodeOutput}
  Gender       CFUP time_to_event group
1 Female -1.6566270           6.1   CosE
2   Male -1.6566270           6.1   CosN
3   Male -0.2051667           6.1   CosG
4   Male  0.3754174           6.1   CosN
5   Male -0.3019307           6.1   CosN
6   Male -0.9792788           6.1   CosJ
\end{CodeOutput}
\end{CodeChunk}

where numerical variables are standardized, and categorical variables with $g$ levels are converted into $(g-1)$ dummy variables.

The aim of applying the time-varying shared frailty Cox model to the \texttt{data$\_$dropout} dataset is to model the time to dropout of students by adjusting for their gender and CFU obtained at the first semester and by considering the students nested structure within degree courses. In particular, the degree course effect is modelled as a time-varying frailty, since we are interested in investigating how the dropout phenomenon does differ across degree courses, across time, net to the effect of the students' characteristics. \\
%

To apply the proposed model, the formula object has to be defined (structured as indicated in Section~\ref{package}). It contains on the left hand side the \texttt{time-to-event} variable and on the right hand side the two regressors \texttt{Gender}, \texttt{CFUP} and the cluster variable \texttt{group}, as follows:
\begin{CodeChunk}
\begin{CodeInput}
R> formula <- time_to_event ~ Gender + CFUP + cluster(group)
\end{CodeInput}
\end{CodeChunk}

The follow-up period starts at $t=1$ and ends at $t=6$ semesters. Indeed, previous studies highlight that drop out happening before the end of the first semester ($t=1$) is very heterogeneous and unpredictable \citep{cannistra2022early, masci2023modelling}, thus not modelled. 
The temporal domain can be divided into intervals in various ways. One straightforward approach is to use the academic structure, dividing the domain into semesters, defining $T_{\text{axis}} = [1.0, 2.0, 3.0, 4.0, 5.0, 6.0]$. Alternatively, intervals can be chosen based on the baseline hazard function's shape, estimated from a time-independent model, e.g., a Shared Gamma Frailty Model. For instance, intervals can be centered around peaks or plateaus of the hazard curve. This latter approach leverages prior evidence for the estimation of the baseline log-hazard $\phi_k$.
Even if this approach requires a further model evaluation, we suggest to follow this procedure since it exploits previous evidence to tailor the shape of our piecewise-linear baseline.
Further details may be found in Appendix \ref{app:example_time_unvarying}.
By following this second approach, the discretized time-domain is chosen to be $T_{\text{axis}} = [1.0, 1.4, 1.8, 2.3, 3.1, 3.8, 4.3, 5.0, 5.5, 5.8, 6.0]$, according to the shape of the baseline hazard function of Figure~\ref{baseline_hazard}.
\begin{CodeChunk}
\begin{CodeInput}
R> time_axis <- c(1.0, 1.4, 1.8, 2.3, 3.1, 3.8, 4.3, 5.0, 5.5, 5.8, 6.0)
\end{CodeInput}
\end{CodeChunk}

%

To execute the main model and optimize the log-likelihood function in a constraint multi-dimensional way, we need to provide a range to each parameter category defined in the vector $\boldsymbol{p}$. We recall that not all parameters are theoretically constrained, but we identify a range to help the optimization procedure to locate the optimum. Since we expect the not random parameters to be only slightly affected by the assumptions on the random part (i.e., the frailty), to define the ranges, we rely on the estimates obtained by a Shared Frailty Model. 
In the following, we report the chosen ranges and specify the motivation behind each one of them:
\begin{itemize}
    \item $\boldsymbol{\phi}$: by looking at the shape of the baseline obtained from a Shared Frailty Model reported in Appendix \ref{app:example_time_unvarying}, 
    we observe that this function is expected to exist in the range $[0,1]$; however, we are going to estimate the baseline log-hazard and, therefore, the argument of the exponential must be negative and located in the range $[-\infty, 0]$. Due to the impossibility of reaching values so close to both boundaries, we set this range to $[-8, -\epsilon]$, where $\epsilon = 1e-10$.
    \item $\boldsymbol{\beta}$: From the application of the Shared Frailty Model reported in Appendix \ref{app:example_time_unvarying}, we get the estimate of the regression coefficients $\texttt{GenderMale}$ and 
    $\texttt{CFUP}$, respectively equal to $0.330^{**}$ and $-1.295^{***}$\footnote{$\text{*}p<0.05; \text{ **}p<0.01; \text{ ***}p<0.001$.}. 
    We therefore set the range for both parameters equal to $[-1.5,0.5]$.
    \item $\mu_1$: since $\mu_1 + \mu_2 = 1$ and $\mu_1, \mu_2 > 0$, a reasonable range for this parameter may be $[\epsilon,1-\epsilon]$.
    \item $\nu$: since we only know that $\nu > 0$ and we do not have any prior information on the right boundary, we set the range equal to $[\epsilon, 1]$. 
    Further insights on the range choices are available in Section \ref{1D_analysis}.
    \item $\boldsymbol{\gamma}$: since we only know that $\gamma > 0$ and we do not have any prior information on the right boundary, we set the range equal to $[\epsilon, 10]$. Further insights on the range choices are available in Section \ref{1D_analysis}.
\end{itemize}
\begin{CodeChunk}
\begin{CodeInput}
R> eps <- 1e-10
R> categories_range_min <- c(-8, -2, eps, eps, eps)
R> categories_range_max <- c(-eps, 0.5, 1 - eps, 1, 10)
\end{CodeInput}
\end{CodeChunk}

\subsection{Model execution and computed parameters}\label{model_execution_output}

\subsubsection{Main model and summary}

Letting unchanged the value of the default variables described in Section \ref{sec:AdPaikModel_fun}, we call the main model with the mentioned input. 

\begin{CodeChunk}
\begin{CodeInput}
R> result <- AdPaikModel(formula, data_dropout, time_axis,
                         categories_range_min, categories_range_max)
\end{CodeInput}
\end{CodeChunk}

To have a comprehensible overview of the output, we exploit our implemented summary method, that receives as unique argument the result of the call:
\begin{CodeChunk}
\begin{CodeInput}
R> summary(result)
\end{CodeInput}
\begin{CodeOutput}
Output of the 'Adapted Paik et al.'s Model'
-------------------------------------------------------------------------------
Call:  time_to_event ~ Gender + CFUP + cluster(group)
with cluster variable ' group ' ( 16 clusters).
-------------------------------------------------------------------------------
Log-likelihood:            -2175.135
AIC:                        4398.2699
Status of the algorithm:    TRUE (Convergence in  31  runs).
-------------------------------------------------------------------------------
Overall number of parameters  24,
divided as (phi, betar, mu1, nu, gammak) = ( 10 , 2 , 1 , 1 , 10 ),
with: number of intervals = 10
      number of regressors = 2 .
-------------------------------------------------------------------------------
Estimated regressors (standard error):
GenderMale : 0.2178 (0.052)
CFUP : -1.2707 (0.0346)
-------------------------------------------------------------------------------
\end{CodeOutput}
\end{CodeChunk}
The summary is synthetic and reports the most important information of the model call: the formula, the cluster variable and its number of levels, the value of the log-likelihood function, the AIC and the status of the algorithm at the end of the optimization procedure. Moreover, it also indicates the number of parameters characterizing the model and how they are subdivided into the different five categories, specifying also their name. For a matter of brevity, only the estimated regression parameters are listed, together with their standard error. 
%
The model requires $27$ iterations (\lq\lq TRUE\rq\rq \, status of the algorithm) 
to reach the convergence on the log-likelihood value.
%

%
\subsubsection{Estimated baseline hazard step-function}
The first category of estimated parameters coincides with the baseline log-hazard $\hat{\boldsymbol{\phi}}$ and it has a numerosity equal to the number of intervals of the temporal domain. \\
The piecewise linear estimated baseline hazard can be obtained 
by applying the exponential operation to $\hat{\boldsymbol{\phi}}$.
These values are reported in the same S3 object and can be retrieved as:
\begin{CodeChunk}
\begin{CodeInput}
R> result$BaselineHazard
\end{CodeInput}
\end{CodeChunk}

To visualize the piecewise linear estimated baseline hazard, we employ \texttt{plot\_bas\_hazard()}, where the x-axis corresponds the effective temporal domain.


\begin{CodeChunk}
\begin{CodeInput}
R> plot_bas_hazard(result)
\end{CodeInput}
\end{CodeChunk}
\begin{figure}[]
    \centering
    \includegraphics[width=0.75\textwidth]{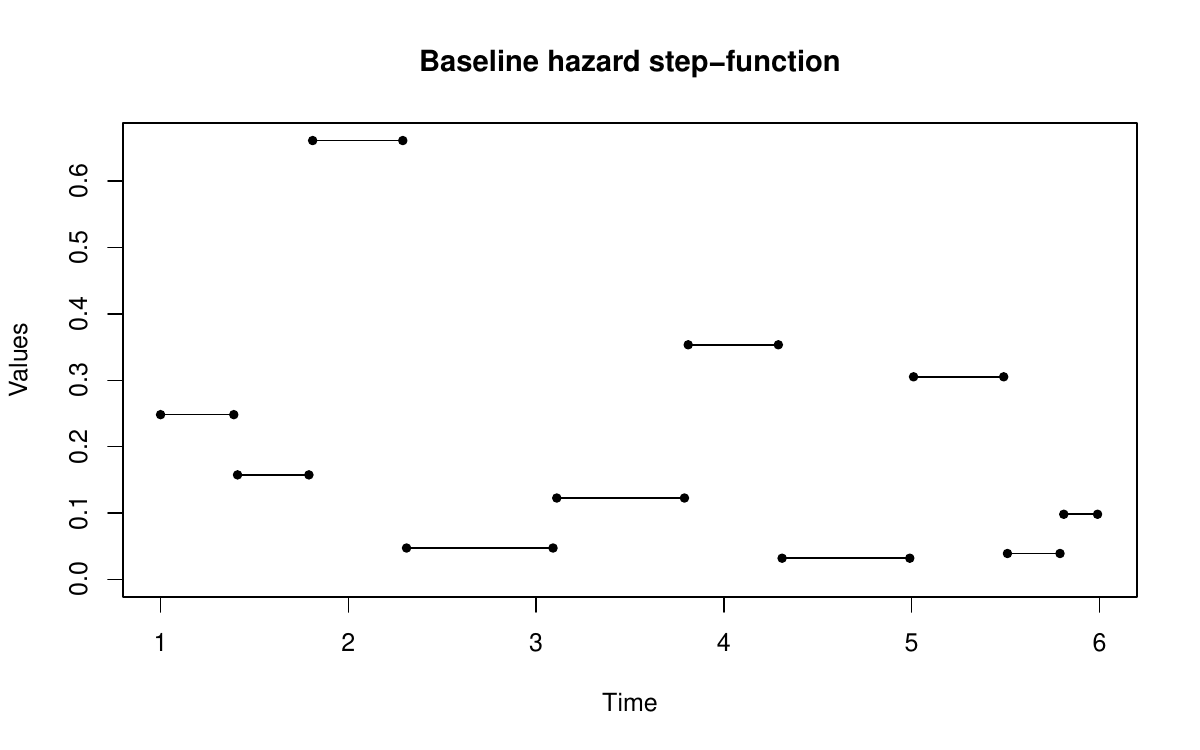}
    \caption{Estimated baseline hazard step-function, Adapted Paik et al.'s Model.}
    \label{baseline_hazard_stepfunction}
\end{figure}
The result is reported in Figure~\ref{baseline_hazard_stepfunction} and shows a correspondence in shape with the curve of Figure~\ref{baseline_hazard}, where results obtained through a time-independent shared frailty Cox model are reported. 
The observed pattern can be explained by the fact that dropout often occurs at the end of the academic year, which accounts for the peaks at times 2 and 4. However, unlike other university datasets, students at this specific university can drop out at any point during the year. This flexibility explains the additional peak at time 5.

\subsubsection{Estimated regressors and frailty parameters}
The second category of estimated parameters $\hat{\boldsymbol{\beta}}$ coincides with the dataset regressors specified in the formula object. 
They are easily available through the \texttt{summary()} function.
As we can observe, they are very close to the ones obtained with the Shared Gamma Frailty Model, in Appendix \ref{app:example_time_unvarying}.
Moreover, all the optimal estimated parameters, divided in each category ($\hat{\boldsymbol{\beta}}$, $\mu$, $\nu$ and $\boldsymbol{\gamma}$), can be easily accessed as  

\begin{CodeChunk}
\begin{CodeInput}
R> coef(result)
\end{CodeInput}
\begin{CodeOutput}
$phi
 [1] -4.077589 -4.532377 -3.098065 -5.734738 -4.782606 
 -3.724189 -6.126564 -3.870919 -5.926393 -5.005396

$beta
GenderMale       CFUP 
  0.217802  -1.270657 

$mu1
[1] 5.99448e-07

$nu
[1] 0.9999995

$gamma
[1] 0.006187576 0.073840907 0.077582404 0.005930964 
0.037877670 0.079786892 0.136952864 0.164864644
\end{CodeOutput}
\end{CodeChunk}

We observe their estimated values are contained in the associated range, with the exception of $\nu$ that assumes a value on the right boundary, which motivation is discussed in Section \ref{1D_analysis}. 

Moreover, their respective standard errors can be obtained as
\begin{CodeChunk}
\begin{CodeInput}
R> coefse(result)
\end{CodeInput}
\begin{CodeOutput}
$se.phi
 [1] 0.1170240 0.1654903 0.1001002 0.2053718 0.1484489 
 0.1238189 0.2970084 0.1582006 0.4088651 0.3170335

$se.beta
GenderMale       CFUP 
0.05201513 0.03457691 

$se.mu1
[1] 0.005022888

$se.nu
[1] 218.6432

$se.gamma
[1] 0.00010000 0.13621989 0.06276072 0.00010000 0.12465050 
0.07305099 0.53064901 0.12265649
\end{CodeOutput}
\end{CodeChunk}
%
%
%

and their respective $95\%$ confidence intervals as
\begin{CodeChunk}
\begin{CodeInput}
R> confint(result)
\end{CodeInput}
\end{CodeChunk}

%

\subsection{Analysis of the frailty}
\label{sec:analysis_of_frailty}

\subsubsection{Frailty standard deviation}
\label{sec:frailtysd}

\begin{figure}[]
    \centering
    \includegraphics[width=0.75\textwidth]{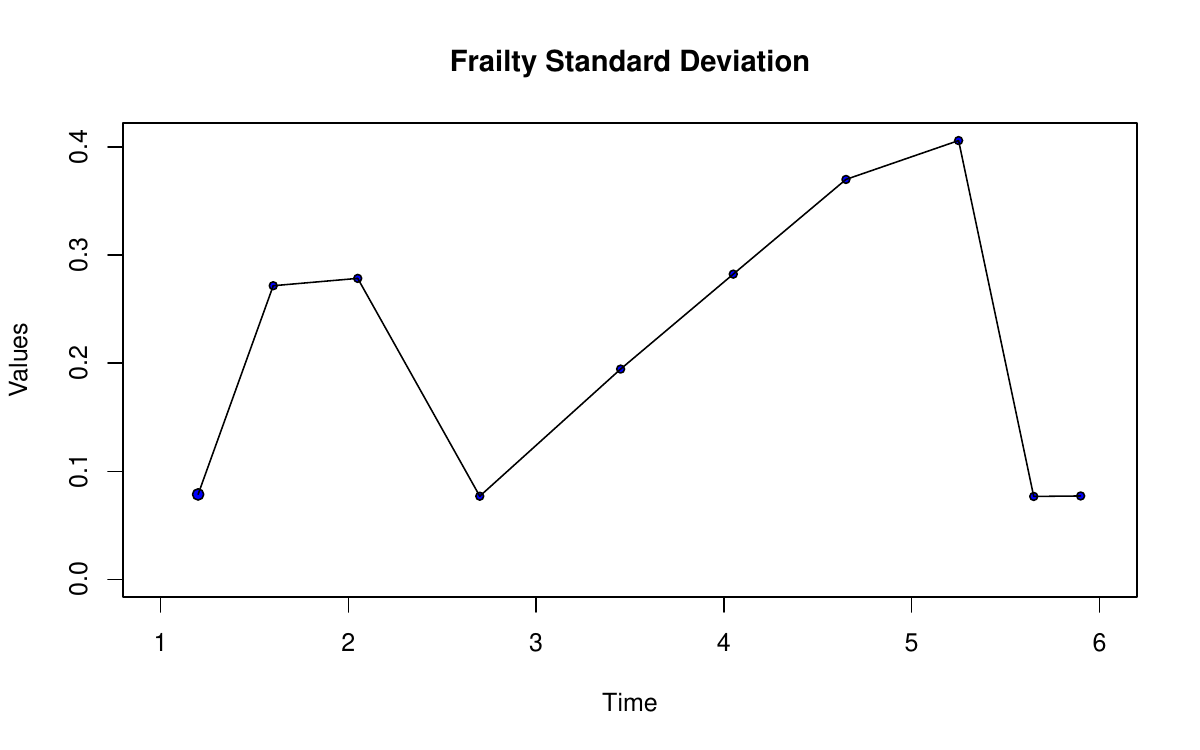}
    \caption{Estimated frailty standard deviation over time.}
    \label{frailty_sd}
\end{figure}

The frailty variance/standard deviation, estimated for each time interval, measure the temporal evolution of the data heterogeneity at the group level. These quantities are computed by the model and stored into an S3 class object called \texttt{FrailtyDispersion}, that can be extracted as:
\begin{CodeChunk}
\begin{CodeInput}
R> result$FrailtyDispersion$FrailtyVariance
R> result$FrailtyDispersion$FrailtyStandardDeviation
\end{CodeInput}
\end{CodeChunk}
To visualize this standard deviation, use the \texttt{plot\_frailty\_sd()} function:
\begin{CodeChunk}
\begin{CodeInput}
R> plot_frailty_sd(result)
\end{CodeInput}
\end{CodeChunk}
The output is shown in Figure~\ref{frailty_sd}.
We observe how the unexplained heterogeneity at the group level evolves in time, confirming the importance of introducing a time-dependent frailty term. This heterogeneity increases in the first three intervals and from the fourth to the eighth, and elsewhere it has almost null value (i.e. $I_4 = [2.3,3.1)$, $I_9 = [5.5, 5.8)$ and $I_{10} = [5.8,6.0]$). \\Being the estimated frailty mean $E[\hat{Z}_{jk}] = \hat{\mu}_1 + \hat{\mu}_2 \approx 1$, a standard deviation close to $0$ in certain intervals implies that the estimated frailty terms are very similar across the groups and not different from the unitary value. Hence, in these time intervals, the dropout phenomenon does not vary across degree courses, net to the effect of students' characteristics. On the other side, a high value of the frailty standard deviation implies higher levels of heterogeneity across degree courses, in terms of dropout risk.
%


On the other hand, the frailty variance can be plotted by:
\begin{CodeChunk}
\begin{CodeInput}
R> plot_frailty_sd(result, flag_variance = TRUE)
\end{CodeInput}
\end{CodeChunk}

As discussed in Section~\ref{adapted_paik_eam}, the frailty variance includes both time-varying and time-unvarying components of heterogeneity. To allow for the analysis of the only time-dependent spread (excluding the constant one), a method is implemented to calculate (and afterwards plot) the reduced frailty standard deviation

\begin{CodeChunk}
\begin{CodeInput}
R> red_frailty_sd <- frailty_sd(result, flag_fullsd = FALSE)
R> plot_frailty_sd(result, frailty_sd = red_frailty_sd, flag_variance = FALSE)
\end{CodeInput}
\end{CodeChunk}
Indeed, the following options in \texttt{plot\_frailty\_sd()} are available: \texttt{frailty\_sd} specifies the standard deviation to plot; if \texttt{flag\_variance} = \lq\lq TRUE\rq\rq \, the variance is plotted instead of the standard deviation; if \texttt{flag\_sd\_external} = \lq\lq TRUE\rq\rq, the standard deviation is passed through \texttt{frailty\_sd}.


\subsubsection{Posterior Frailty Estimates}
\label{sec:postfrailtyest}

Similarly, we analyze the posterior frailty estimates computed by the model and saved under 

\texttt{result\$PosteriorFrailtyEstimates}, to study firstly the impact of each degree course on the instantaneous risk of withdrawal university and then its evolution in time. Such an analysis is crucial for identifying time-dependent differences across degree courses, which can inform targeted interventions. \\
The posterior estimates are available for each component of the frailty structure (i.e., $\alpha_j$, $\epsilon_{jk}$, $\forall j,k$) as well as the full term ($Z_{jk}$), which is normalized to ensure a unitary frailty mean. While these estimated matrices can be easily accessed, using the implemented plot method provides a clearer understanding of their behavior and risk impact. \\
To distinguish the temporal estimates of each degree course, we use different colors and point shapes, resulting in 16 unique combinations. To plot the posterior full frailty estimates $\hat{Z}_{jk}$, we set both the arguments \texttt{flag$\_$eps} and \texttt{flag$\_$alpha} equal to \lq\lq FALSE\rq\rq (default values) as follows:

\begin{CodeChunk}
\begin{CodeInput}
R> pch_type <- c(21, seq(21,25,1), seq(21,25,1), seq(21,25,1))
R> color_bg <- c("darkblue", rep("red", 5), rep("purple", 5), rep("green",5))
R> plot_post_frailty_est(result, data_dropout,
                      pch_type = pch_type, color_bg = color_bg)
\end{CodeInput}
\end{CodeChunk}
\begin{figure}[]
    \centering
    \includegraphics[width=0.75\textwidth]{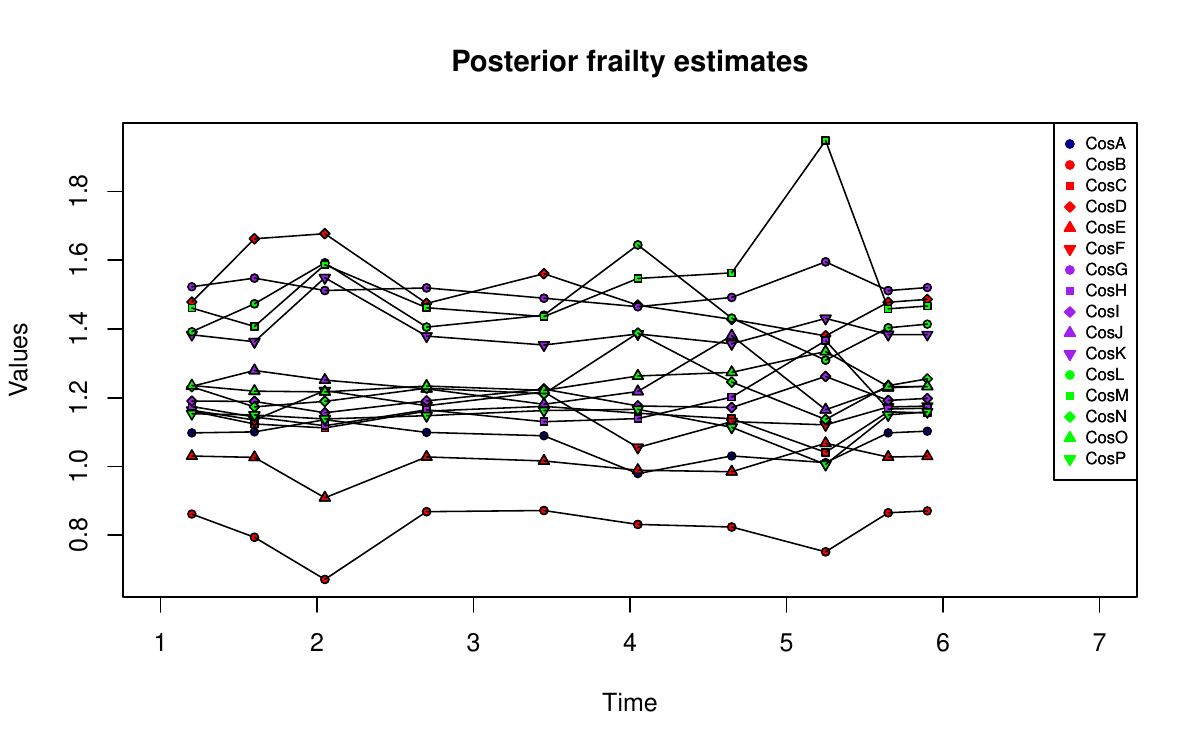}
    \caption{Posterior frailty estimates of $\hat{Z}_{jk}$ over time.}
    \label{full_post_frailty_est}
\end{figure}

Figure~\ref{full_post_frailty_est} illustrates how the posterior estimates for each degree course evolve over time. The spread of the piecewise linear curves in certain intervals aligns with the estimated time-varying frailty standard deviation.
The estimated posterior mean obtained as
\begin{CodeChunk}
\begin{CodeInput}
R> mean(result$PosteriorFrailtyEstimates$Z)
\end{CodeInput}
\begin{CodeOutput}
[1] 1.239546
\end{CodeOutput}
\end{CodeChunk}
suggests that most faculties have posterior frailties that slightly increase the instantaneous dropout risk across all intervals, with the course $CosM$ nearly doubling the risk in the eighth interval. Conversely, $CosB$ consistently shows estimates below the unitary value, indicating a lower risk of withdrawal for its students throughout the follow-up period.

\begin{figure}[]
    \centering
    \includegraphics[width=0.45\textwidth]{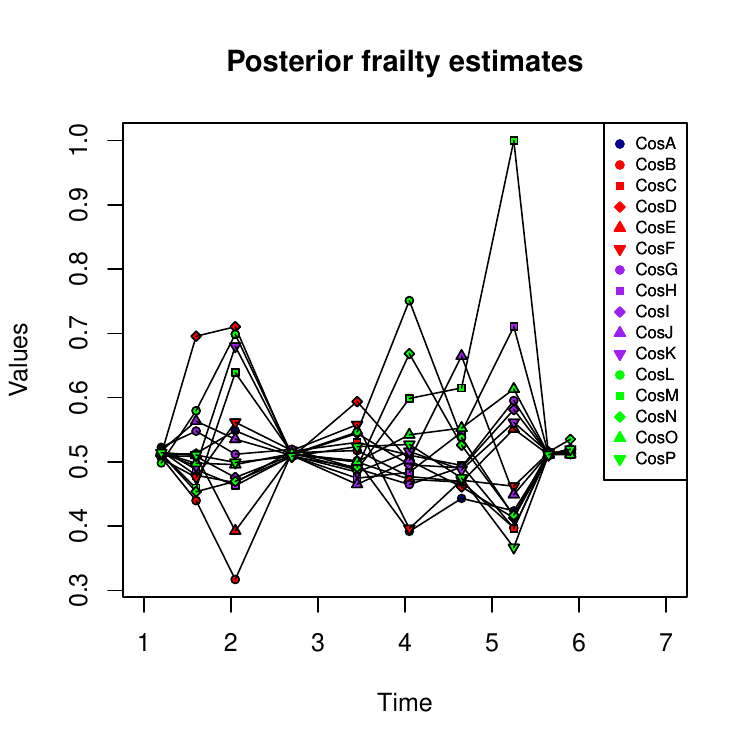}
    \includegraphics[width=0.45\textwidth]{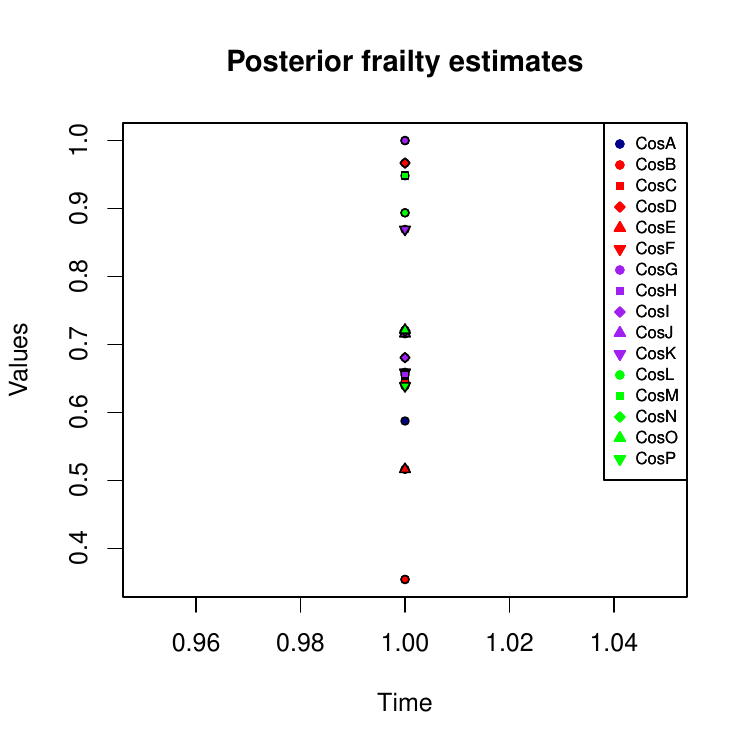}
    \caption{Posterior frailty estimates of $\hat{\epsilon}_{jk}$ (left panel) and $\hat{\alpha}_{j}$ (right panel).}
    \label{eps_post_frailty_est}
\end{figure}

This result allows us to identify, for each degree course, the time instants and the semesters in which the students are more at risk of dropout. 
\\
In addition to visualize the full frailty $\hat{Z}_{jk}$, it is possible to plot only the time-varying component $\hat{\epsilon}_{jk}$ or the constant term $\alpha_{j}$ by setting the \texttt{flag\_eps} and \texttt{flag\_alpha} equal to \lq\lq TRUE\rq\rq. When plotting the constant term, the x-axis scale is adjusted since only one point per degree course is plotted. The associated code is shown below and the resulting plot is reported in Figure~\ref{eps_post_frailty_est}.

\begin{CodeChunk}
\begin{CodeInput}
R> plot_post_frailty_est(result, data_dropout,
                      flag_eps = TRUE, flag_alpha = FALSE,
                      pch_type = pch_type, color_bg = color_bg)
\end{CodeInput}
\end{CodeChunk}

\begin{CodeChunk}
\begin{CodeInput}
R> plot_post_frailty_est(result, data_dropout,
                      flag_eps = FALSE, flag_alpha = TRUE,
                      pch_type = pch_type, color_bg = color_bg)
\end{CodeInput}
\end{CodeChunk}
%

Comparing the left panel of Figure~\ref{eps_post_frailty_est} with Figure~\ref{frailty_sd} further confirms the presence of specific time intervals with high variability across degree courses.

%

%
%

\subsection{Conditional Survival Function}
\label{sec:condsurvfun}

Finally, the conditional survival function described in Eq. (\ref{eq:condsurvfun}) can be computed and plotted as follows.

\begin{CodeChunk}
\begin{CodeInput}
R> survival_df = survival(result, data_dropout)
R> plot_survival(result, survival_df)
\end{CodeInput}
\end{CodeChunk}

\begin{figure}[]
    \centering
    \includegraphics[width=0.75\textwidth]{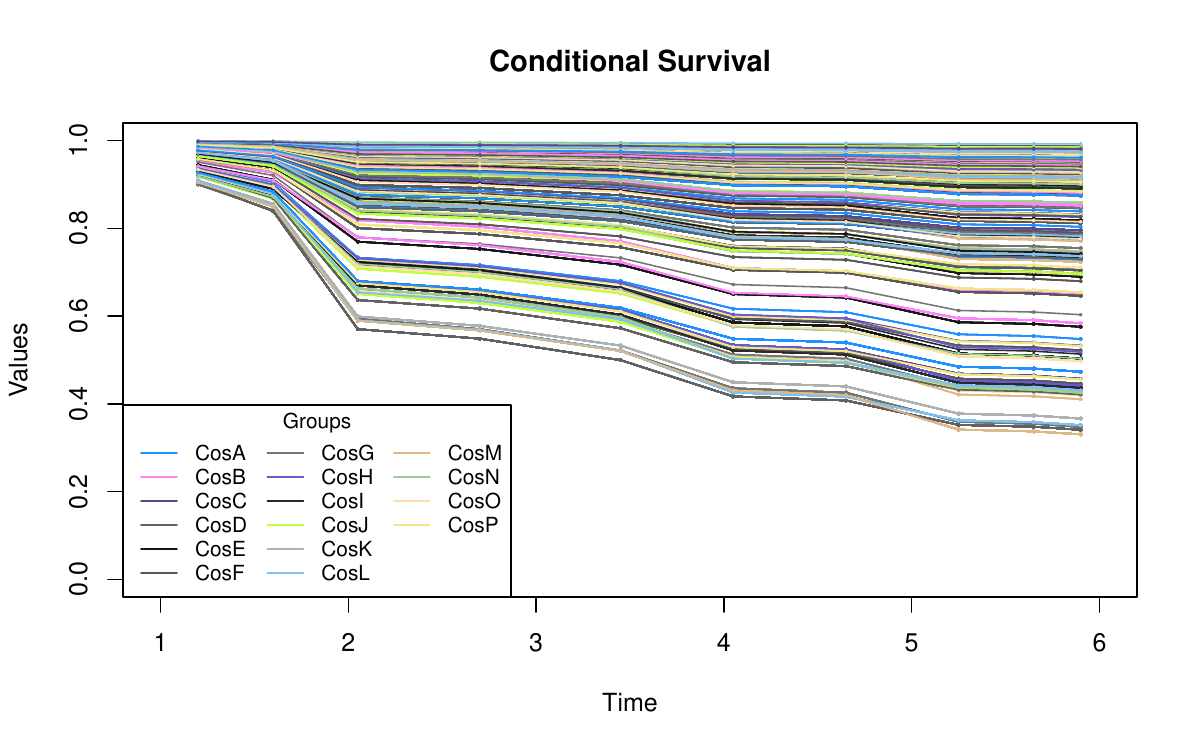}
    \caption{Conditional Survival function for each patient.}
    \label{survival}
\end{figure}

In Figure \ref{survival}, one curve per individual is plotted, with a color related to the group the unit belongs.

\subsection{1D analysis of the log-likelihood}
\label{1D_analysis}


In this subsection, we provide a practical example of using \texttt{AdPaik\_1D()} on the \texttt{data\_dropout} dataset. The following two subsections will separately describe two distinct uses of the function.

\subsubsection{Identify a parameter existence range}
\label{sec:Identify a parameter existence range}
\begin{figure}[]
  \subfloat[$\phi_1$, parameter 1]{
	\begin{minipage}[c][1\width]{
	   0.45\textwidth}
	   \centering
	   \includegraphics[width=1\textwidth,page=5]{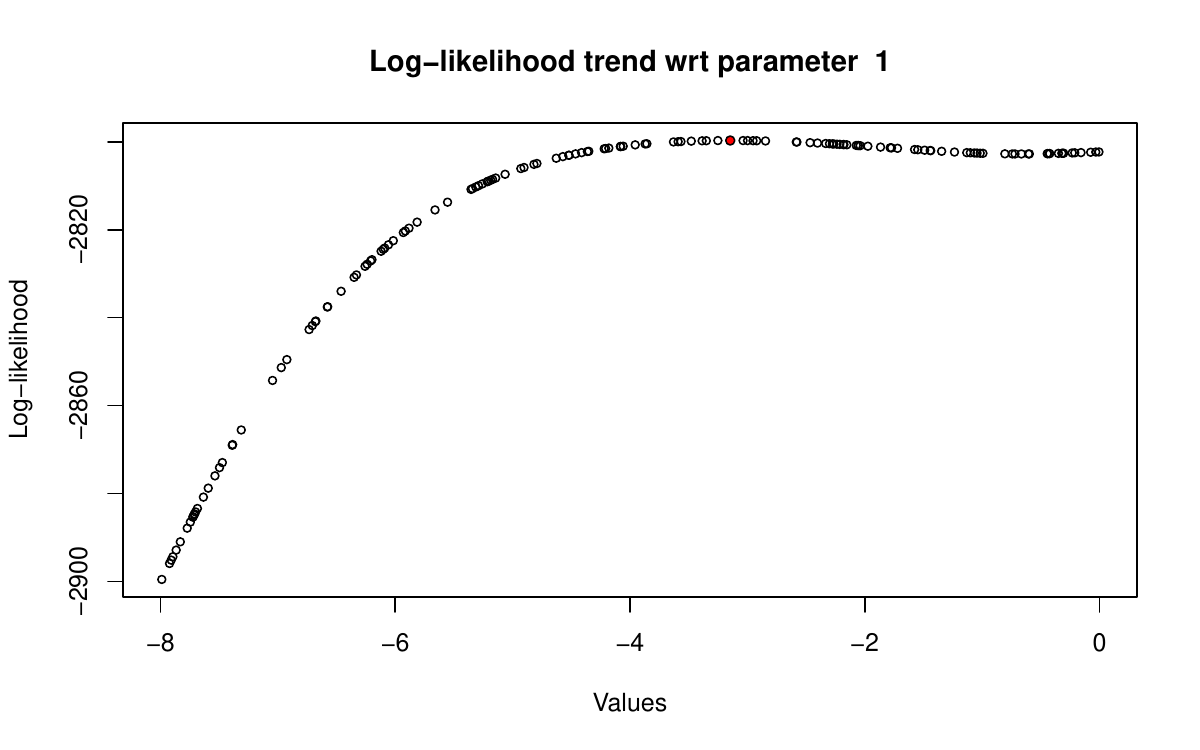}
     \vspace{-3cm}
        \label{trend_ll_phi1_1}
	\end{minipage}}
 \hfill 	
  \subfloat[$\mu$, parameter 12]{
	\begin{minipage}[c][1\width]{
	   0.45\textwidth}
	   \centering
	   \includegraphics[width=1\textwidth,page=5]{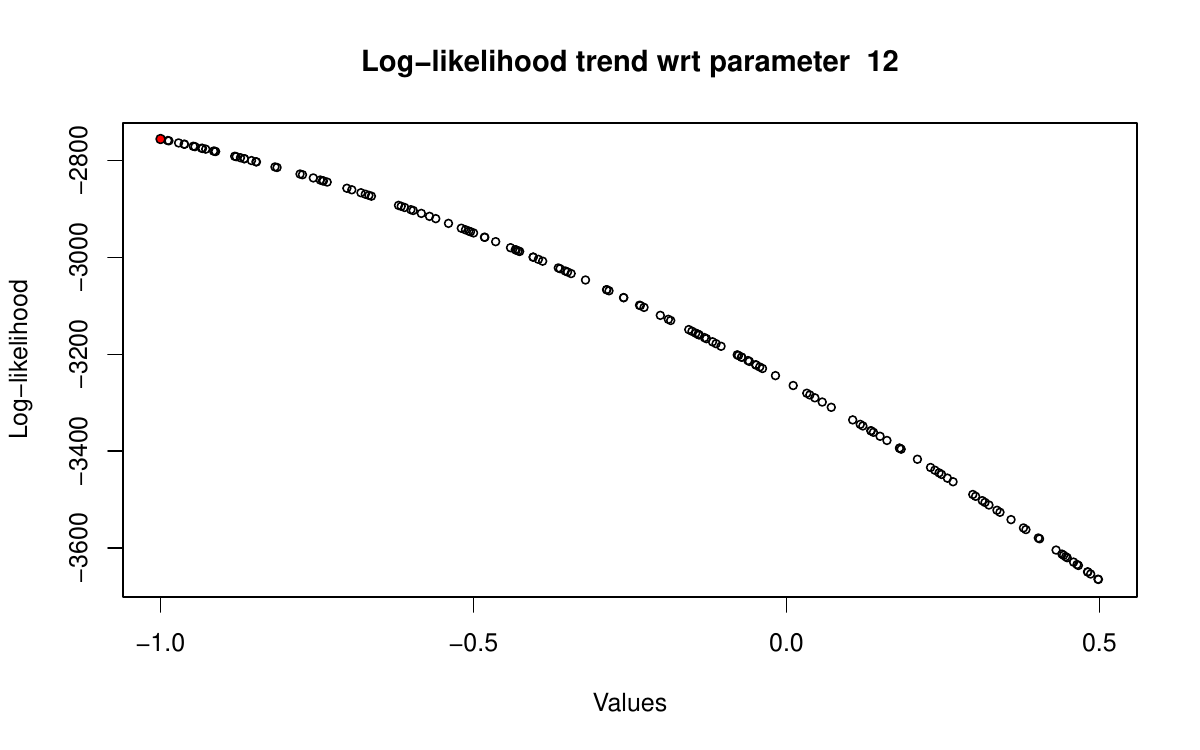}
     \vspace{-3cm}
    \label{trend_ll_beta1_1}
	\end{minipage}}
\caption[Trend of the log-likelihood function with respect to two parameters]{Trend of the log-likelihood function with respect to two parameters.}
\end{figure}

\texttt{AdPaik\_1D()} can be used to identify a possible parameters range. 
In fact, the algorithm makes use of a constrained optimization procedure that requires both the minimum and the maximum of the parameter's range. This function helps to carefully choose them in two ways: first, it narrows the potentially vast search space in $R^{n_p}$ to a smaller, more suitable interval for the parameter of interest; second, it enforces any required existence condition. 
%
%
This step should be performed before running the main model and repeated \textit{n\_iter} times for each parameter.
\\\\
For example, if we analyze parameter $\phi_1$ in position index $1$, and we set \textit{n$\_$iter = 5} and all the other parameters are let to assume a random value in the ranges chosen before, we obtain 
%
\begin{CodeChunk}
\begin{CodeInput}
R> index_param_to_vary <- 1
R> analysis_1D_opt <- AdPaik_1D(formula, data_dropout,
                                time_axis, index_param_to_vary, 
                                flag_optimal_params = FALSE, 
                                optimal_params = NULL,
                                categories_range_min, categories_range_max, 
                                n_iter = 5)
R> analysis_1D_opt
\end{CodeInput}
\begin{CodeOutput}
$EstimatedParameter
[1] -2.548404 -1.875524 -1.487164 -3.125632 -3.146015

$OptimizedLoglikelihood
[1] -2601.455 -3726.909 -2897.432 -3029.110 -2799.696

attr(,"class")
[1] "AdPaik_1D"
\end{CodeOutput}
\end{CodeChunk}
Agreeing with the imposed range [$-8, -\epsilon$].
An implicit utility derived from plotting the trend of the log-likelihood function (setting \texttt{flag\_plot} = \lq\lq TRUE\rq\rq) consists in checking that the estimated parameter value actually coincides with a likelihood's maximum.
As we can observe in Figure~\ref{trend_ll_phi1_1}, the function increases up to the maximum point and then decreases.\\
As a second example, if we consider the parameter $\beta_2$, with index $12$, and we change the range of the regressors category as follows:
\begin{CodeChunk}
\begin{CodeInput}
R> index_param_to_vary <- 12
R> categories_range_min <- c(-8, -1, eps, eps, eps)
R> categories_range_max <- c(-eps, 0.5, 1 - eps, 1, 10)
\end{CodeInput}
\end{CodeChunk}
we would fall into a situation where the initial provided parameter range does not include the real value (because $\beta_2 < [-1, 0.5]$). The output reports
\begin{CodeChunk}
\begin{CodeOutput}
$EstimatedParameter
[1] -0.9999995 -0.9999995 -0.9999995 -0.9999995 -0.9999995

$OptimizedLoglikelihood
[1] -2701.092 -2783.330 -2684.514 -2566.700 -2755.624

attr(,"class")
[1] "AdPaik_1D"
\end{CodeOutput}
\end{CodeChunk}
and the trend of the log-likelihood function is reported in Figure~\ref{trend_ll_beta1_1}. These two aspects warn us about the non-correctness of the provided range,
implying that the maximum point is likely reached in a different interval.
%
%

\subsubsection{Study the log-likelihood behaviour}
The other use of \texttt{AdPaik\_1D()} is the analysis of the log-likelihood function.
Specifically, we employ this function to study the estimated parameter $\hat{\nu}$, which is the only parameter that reaches the upper boundary of the provided range. 

\begin{figure}[]
    \centering
    \includegraphics[width=0.75\textwidth]{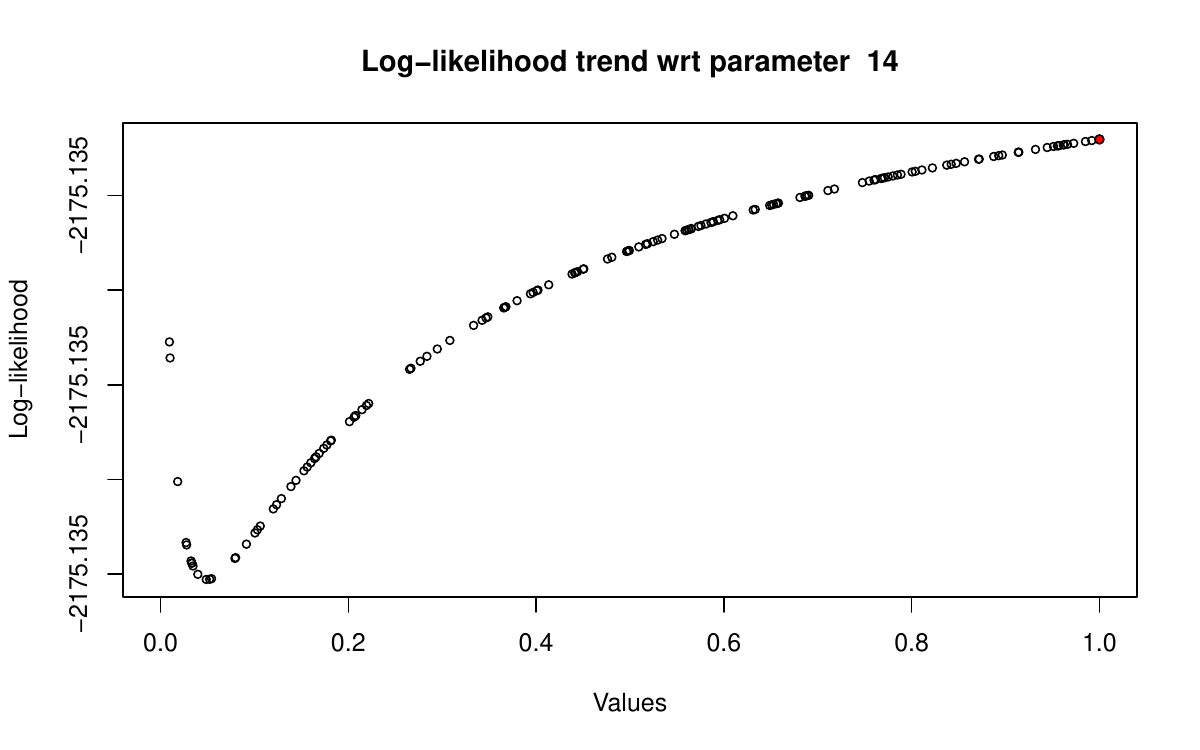}
    \caption{Trend of the log-likelihood function, with respect to the parameter $\nu$.}
    \label{trend_ll_14}
\end{figure}

For this analysis, we utilize the optimal parameter values from the main model, setting the following flags and parameters: \texttt{flag\_optimal\_params} = \lq\lq TRUE\rq\rq, \texttt{optimal\_params = result\$OptimalParameters}, and \texttt{n\_iter = 1}. The \texttt{n\_iter} parameter, which controls how many iterations to perform, is set to 1 because we aim to optimize the log-likelihood with respect to $\nu$, while keeping all other parameters fixed at their optimal values. 
Moreover, since the parameter $\nu$ is indexed at position 14 in the parameter vector $\boldsymbol{p}$, we set 
\begin{CodeChunk}
\begin{CodeInput}
R> categories_range_min <- c(-8, -2, eps, eps, eps)
R> categories_range_max <- c(-eps, 0.4, 1 - eps, 1, 10)
R> index_param_to_vary <- 14
R> analysis_1D_opt <- AdPaik_1D(formula, 
                    data_dropout, 
                    time_axis,
                    index_param_to_vary, 
                    categories_range_min, 
                    categories_range_max, 
                    flag_optimal_params = TRUE,
                    flag_plot = TRUE,
                    optimal_params = result$OptimalParameters,
                    n_iter = 1)
                    
R> analysis_1D_opt
\end{CodeInput}
\begin{CodeOutput}
$EstimatedParameter
[1] 0.9999995

$OptimizedLoglikelihood
[1] -2175.135

attr(,"class")
[1] "AdPaik_1D"
\end{CodeOutput}
\end{CodeChunk}
The output \texttt{analysis\_1D\_opt} provides both the estimated optimal value which maximizes the log-likelihood function, and the corresponding log-likelihood value. Notably, this value matches the one obtained from the main model\footnote{\texttt{result\$OptimalParameters[index\_param\_to\_vary]}}.
What stands out in this analysis is the generated plot, displayed in Figure~\ref{trend_ll_14}. As we can see, the log-likelihood function initially decreases near 0 and then increases toward the upper boundary of the given interval. Consequently, the maximum value occurs close to 1.
However, if we focus on the y-axis, we notice that the log-likelihood is almost constant across the entire range, and the optimal value corresponds to the produced highest point.
%
%
From the main model's output, we observe that the estimated parameter $\hat{\nu}$ consistently reaches the upper boundary of the provided range, unlike other parameters. 
However, the upper boundary of the provided range is also reached when the range is changed.
For example, changing the range for $\nu$ from $[\epsilon, 1]$ to $[\epsilon, 2]$ still results in $\hat{\nu}$ aligning with the upper limit and in the same optimized log-likelihood value:
\begin{itemize}
    \item Range of $\nu$: $[\epsilon, 1]$ $\to$ $\hat{\nu} = 0.9999995$\\
    Optimal log-likelihood value = $-2175.135$
    \item Range of $\nu$: $[\epsilon, 2]$ $\to$ $\hat{\nu} = 1.999999$\\
    Optimal log-likelihood value = $-2175.135$
\end{itemize}

These results suggest the independence of the log-likelihood function with respect to the parameter $\nu$, as also indicated by the important value of its standard error (obtained through 

\texttt{result\$StandardErrorParameters[L+R+2]}), estimated equal to $218.643$.
However, this unusual behavior seems to be specific to the dataset used and may not generalize to other datasets. Different structures or levels of heterogeneity in other datasets could yield different patterns. In our case, this effect may stem from significant data heterogeneity within the time-independent scenario, as modeled using the Time-Independent Shared Gamma-Frailty Cox Model.

\section{Summary and Discussion}\label{Conclusion}

\pkg{TimeDepFrail} is an \proglang{R} package designed to implement time-varying shared frailty models.
By partitioning the analysis domain into multiple time intervals, this package allows frailty terms to vary both across groups and over time, offering a versatile and flexible framework for addressing time-dependent heterogeneity in the analysis of survival data. \\\\
We believe that \pkg{TimeDepFrail} will be particularly valuable for practitioners who need to fit shared frailty models in contexts where the assumption of constant heterogeneity over time might be too restrictive. The package simplifies the use of these models, making them more accessible to a wider audience, setting the basis for further advancements in time-varying frailty modelling. \\\\
\pkg{TimeDepFrail} offers a complete set of functions for model fitting and results interpretation and visualization, however, it is not without its limitations. 
Currently, the implemented method is tailored to a specific form of time-dependent frailty with a pre-defined structure, which may not offer complete flexibility for all the modeling needs\footnote{Other time-dependent frailty models are discussed in \citep{wintrebert2004centre}, but they suffer the lack of an efficient \cite{R} implementation.}.
Additionally, the choice of the initial parameters ranges plays an important role in the results, requiring careful consideration. The knowledge and the application of a classical time-independent Cox model helps out in this identification phase, but it may not be sufficient, notably for the parameters related to the frailty terms.\\
Alternative approaches, such as those based on Lévy processes, as implemented in \pkg{dynfrail} \citep{dynfrail}, offer more theoretical flexibility for modeling time-varying frailty. However, these methods are computationally intensive and less practical for real-world applications due to their complexity and the limited amount of data they can efficiently process. In contrast, \pkg{TimeDepFrail} offers a more practical and computationally manageable option that works well in applied scenarios, making it a valuable tool for practitioners.
Future development of the package could explore the incorporation of other frailty forms mentioned earlier in the paper, further enhancing its adaptability and improving the optimization method, by preferring a more sophisticated one-dimensional search of the optimum. We aim to continue improving \pkg{TimeDepFrail} by extending its capabilities and making it an even more comprehensive tool for survival analysis. \\\\
The package is currently available on GitHub\footnote{At \href{https://github.com/alessandragni/TimeDepFrail}{https://github.com/alessandragni/TimeDepFrail}.} and on CRAN 
\citep{timedepfrail2025}.
We welcome contributions to the package's development and encourage collaboration via pull requests to the GitHub repository.

\section*{Competing interests}
No competing interest is declared.

\begin{appendix}
    \renewcommand\thefigure{\thesection.\arabic{figure}}

\section{Time-Independent Shared Gamma-Frailty Cox Model}
\label{app:Time-Independent Shared Gamma-Frailty Cox Model}

Let $Z_j \sim Gamma(\xi, \xi)$, $\forall j = 1, \dots, N$, where the two parameters are the same because of identifiability issues (the expectation of the frailty is fixed to 1).
The variance of the frailty $Var(Z_j) = 1/\xi = \theta$ quantifies the degree of between-groups variability. 
%
%
In this specific case, it is possible to derive the posterior distribution of the frailty given the observed data within each group, resulting again in a $Gamma$ with different parameters. 
In detail, let $N_j(\tau)$ be the number of observed event in group $j$ up to a certain time-horizon $\tau$ and define the event history of this units as $F_{j,\tau}$. The sum of the cumulative hazard function for the units within group $j$ can be computed as
$H_{j,\bullet}(\tau) = \sum_{i=1}^{n_j} \int_{0}^{\tau} Y_{ij}(s) h_{ij}(s)ds$,
where $Y_{ij}(s)$ is the at risk indicator of subject $i$ in group $j$, equal to $1$ if this subject is at risk at time $s$ and $0$ otherwise. The posterior mean and variance for the frailty are:
\begin{equation*} 
    E[Z_j|F_{j,\tau}] = \frac{\xi + N_{j}(\tau)}{\xi + H_{j,\bullet}(\tau)} \qquad var[Z_j|F_{j,\tau}] = \frac{\xi + N_{j}(\tau)}{(\xi + H_{j,\bullet}(\tau))^2}.
\end{equation*}
%
The \textit{empirical Bayes estimate} $\hat{Z}_j$ for each group $j$ 
can be computed as $\hat{Z}_j = (\hat{\xi} + N_{j}) / (\hat{\xi} + \hat{H}_{j,\bullet})$.
If $\hat{Z}_j<1$, units belonging to group $j$ have a smaller risk of failure with respect to units belonging to an \textit{average} group ($Z_j=1$), with a decrease of $(1-\hat{Z}_j)\%$ of the hazard function. On the other hand, if $\hat{Z}_j>1$, we register an increase in the failure rate meaning that these units have more chance of facing the event with respect to the average.

\section{Proof of loglikelihood function}\label{proof_loglikleihood_function}

We report here the entire proof of the log-likelihood function of the \textit{Adapted Paik et al.'s Model} reported in \cite{wintrebert2004centre}, starting from scratch. \\
%
Let $N$ be the number of groups in which the whole population is divided based on a stratification covariate\footnote{Corresponding to \texttt{group} in our application.}.
\\Let $t_{ij}$ be the time-to-event for the $j$th unit in the $i$th group and let the time-domain be divided into $L$ intervals $I_k = [a_{k-1},a_{k})$, $k = 1, \dots, L$, with discrete time-points $0=a_0 < a_1 < \dots < a_L = \infty$. Moreover, let $d_{ijk}$ be the event variable for the $j$th unit in the $i$th group in $I_k$, such that $d_{ijk} = 1$ if $t_{ij}$ is uncensored in $I_{k}$ and 0 otherwise. The binary indicator $d_{ij}$ of event can be retrieved by $\sum_{k}{d_{ijk}}$ \citep{wintrebert2004centre}.\\\\
Let $Z_{jk}$ be the unobservable frailty of the $j$th group for the $k$th time-interval. Conditionally on $Z_{jk}$, the hazard function $h_{ijk}$ for the $i$th unit, in the $j$th group in $I_k$, is given by the general expression \citep{wintrebert2004centre}:
\begin{equation*}
    h_{ijk}(t_{ij} | Z_{jk}) = Z_{jk} \text{ e}^{(\boldsymbol{\beta}^T \boldsymbol{x}_{ij} + \phi_k)} 
\end{equation*}
Thanks to the further assumption that subjects in the same group are independent, the likelihood $L_j(t|Z_{j.})$ conditionally on the group frailty term $Z_{j.}$ can be obtained as:
\begin{align*}
    & L_j(t|Z_{j.}) = \prod_{i,k} h_{ijk}(t_{ij} | Z_{jk})^{d_{ijk}} \text{ e}^{-H_{ijk}(t_{ij} | Z_{jk})} \\
    & H_{ijk}(t_{ij} | Z_{jk}) = \int_{0}^{t_{ij}} h_{ijk}(s | Z_{jk}) ds 
\end{align*}
where $H_{ijk}(t_{ij} | Z_{jk})$ is the conditional cumulative hazard function at $t_{ij}$. Unless it is differently declared, the latter integral is resolved introducing the following variable
\begin{equation}\label{formula_e_ijk}
    e_{ijk} =\begin{cases}  0 & \text{if } t_{ij} < a_{k-1} \\ t_{ij}-a_{k-1} & \text{if }t_{ij} \in I_k \\ a_{k}-a_{k-1} & \text{if }t_{ij} \geq a_k 
\end{cases}
\end{equation}
so that $H_{ijk}(t_{ij} | Z_{jk}) = e_{ijk} h_{ijk}(s | Z_{jk})$. \\\\
Eventually, if $g(Z_{j.})$ is a general density function for the frailty $Z_{j.}$, then the likelihood $L_j$  can be derived as:
\begin{equation*}
    L_j = \int L_j(t|Z_{j.}) g(Z_{j.})dZ_{j.}
\end{equation*}
Before introducing the frailty term and derive the log-likelihood for the \textit{Adapted Paik et al.'s Model}, we recall the density function for a gamma distributed random variable $x$. \\If $x \sim Gamma(\zeta, \psi)$ with $\zeta, \psi > 0$, then its density function $g(x)$ is:
\begin{equation} \label{gamma_density}
    g(x) = \frac{\psi^{\zeta}}{\Gamma(\zeta)} x^{\zeta - 1} \text{ e}^{- \psi x} \quad \text{ with } \quad \Gamma(\zeta) = \int_{0}^{\infty} x^{\zeta-1} \text{ e}^{-x} dx
\end{equation}\\
In \cite{wintrebert2004centre}, the frailty term $Z_{jk}(t_{ij})$ is constructed to be equal to $Z_{jk}(t_{ij}) = (\alpha_{j} + \epsilon_{jk})$ for $t_{ij} \in I_k$, leading to the hazard function $h_{ijk}(t_{ij} | \alpha_j, \epsilon_{jk}) = (\alpha_{j} + \epsilon_{jk}) \text{ e}^{(\boldsymbol{\beta}^T \boldsymbol{x}_{ij}  + \phi_k)}$, where the parameters $\alpha_j$ and $\epsilon_{jk}$ are chosen to be independent and distributed according to:
\begin{itemize}
    \item $\alpha_j \sim Gamma(\mu_1/\nu, 1/\nu)$ $\forall j$ 
    \item $\epsilon_{jk} \sim Gamma(\mu_2/\gamma_k, 1/\gamma_k)$ $\forall j,k$
\end{itemize}
with $\mu_1, \mu_2,  \nu, \gamma_k > 0$, $\forall k$, and the constraint $E[Z_{jk}] = \mu_1 + \mu_2 = 1$ needed for identifiability.\\
The likelihood $L_j$ for group $j$ is built as follows:
\begin{align*}
    L_j &= \int_{0}^{\infty} \int_{0}^{\infty}  \prod_{i,k}  h_{ijk}(t_{ij} | \alpha_j, \epsilon_{jk})^{d_{ijk}} \text{ e}^{-H_{ijk}(t_{ij} | Z_{jk})} g(\alpha_j) \prod_{k} g(\epsilon_{jk}) d\alpha_j d\epsilon_{jk} \\  &= \int_{0}^{\infty} \int_{0}^{\infty}  \prod_{i,k} [(\alpha_j + \epsilon_{jk}) \text {e}^{(\boldsymbol{\beta}^T \boldsymbol{x}_{ij}  + \phi_k)}]^{d_{ijk}} \text{ e}^{-e_{ijk}(\alpha_j + \epsilon_{jk}) \text{ e}^{(\boldsymbol{\beta}^T \boldsymbol{x}_{ij} + \phi_k)}} g(\alpha_j) \prod_{k} g(\epsilon_{jk}) d\alpha_j d\epsilon_{jk} \\
    &= \prod_{i,k}[\text{e}^{(\boldsymbol{\beta}^T \boldsymbol{x}_{ij}  + \phi_k)}]^{d_{ijk}} \int_{0}^{\infty} \int_{0}^{\infty} \prod_{i,k}(\alpha_j + \epsilon_{jk})^{d_{ijk}} \text{ e}^{-e_{ijk}(\alpha_j + \epsilon_{jk}) \text{ e}^{(\boldsymbol{\beta}^T \boldsymbol{x}_{ij}  + \phi_k)}} g(\alpha_j) \prod_{k} g(\epsilon_{jk}) d\alpha_j d\epsilon_{jk}
\end{align*}
Thanks to the substitution:
\begin{equation*}
    \prod_{i,k}(\alpha_j + \epsilon_{jk})^{d_{ijk}} = \prod_{k} (\alpha_j + \epsilon_{jk})^{d_{j.k}} = \prod_{k} \sum_{l=0}^{d_{j.k}} \alpha_j^l \epsilon_{jk}^{(d_{j.k}-l)} \binom{d_{j.k}}{l} 
\end{equation*}
where $d_{j.k}=\sum_{i}d_{ijk}$, we obtain the following expression for $L_j$: 
\begin{align*}
    L_j &= \prod_{i,k}[\text{e}^{(\boldsymbol{\beta}^T \boldsymbol{x}_{ij}  + \phi_k)}]^{d_{ijk}}  \prod_{k} \sum_{l=0}^{d_{j.k}} \binom{d_{j.k}}{l} \cdot \\
    & \quad \cdot \underbrace{ \int_{0}^{\infty} \prod_{i,k} g(\alpha_j) \alpha_j^l\text{ e}^{-\alpha_j e_{ijk} \text{ e}^{(\boldsymbol{\beta}^T \boldsymbol{x}_{ij}  + \phi_k)}} d\alpha_j}_{(a)} \quad \underbrace{ \int_{0}^{\infty} \prod_{i,k} g(\epsilon_{jk}) \text{ e}^{-\epsilon_{jk} e_{ijk} \text{ e}^{(\boldsymbol{\beta}^T \boldsymbol{x}_{ij}  + \phi_k)}} \epsilon_{jk}^{(d_{j.k}-l)}\prod_{k}d\epsilon_{jk}}_{(b)}  
\end{align*}
where we collect in $(a)$ all the $\alpha_j$ terms and in $(b)$ all the $\epsilon_{jk}$ terms, so that it is possible to analytically solve the integrals using the definition of the gamma function $\Gamma(\alpha)$ of (\ref{gamma_density}). 
\begin{align*}
    (a) &= \int_{0}^{\infty} \prod_{i,k} g(\alpha_j) \alpha_j^l \text{ e}^{-\alpha_j e_{ijk} \text{ e}^{(\boldsymbol{\beta}^T \boldsymbol{x}_{ij}  + \phi_k)}} d\alpha_j = \int_{0}^{\infty} g(\alpha_j) \alpha_j^l \text{ e}^{-\alpha_j A_{j..}} d\alpha_j \\
    &= \int_{0}^{\infty} \frac{\text{e}^{-\alpha_j/\nu}(\alpha_j/\nu)^{\mu_1/\nu}}{\alpha_j \Gamma(\mu_1/\nu)} \alpha_j^l \text{ e}^{-\alpha_j A_{j..}} d\alpha_j =  \frac{\Gamma(\mu_1/\nu+l)}{\Gamma(\mu_1/\nu)} \frac{(1/\nu)^{\mu_1/\nu}}{(1/\nu+A_{j..})^{(\mu_1/\nu+l)}}. \\\\
    (b) &= \int_{0}^{\infty} \prod_{i,k} g(\epsilon_{jk}) \text{ e}^{-\epsilon_{jk} e_{ijk} \text{ e}^{(\boldsymbol{\beta}^T \boldsymbol{x}_{ij}  + \phi_k)}} \epsilon_{jk}^{(d_{j.k}-l)}\prod_{k}d\epsilon_{jk} = \int_{0}^{\infty} \prod_{k} g(\epsilon_{jk}) \epsilon_{jk}^{(d_{j.k}-l)} \text{ e}^{-\epsilon_{jk}A_{j.k}} \prod_{k}d\epsilon_{jk} \\
    &= \int_{0}^{\infty} \prod_{k} \frac{\text{e}^{-\epsilon_{jk}/\gamma_k} \epsilon_{jk}^{\mu_2/\gamma_k-1}(1/\gamma_k)^{\mu_2/\gamma_k}}{\Gamma(\mu_2/\gamma_k)} \epsilon_{jk}^{(d_{j.k}-l)} \text{ e}^{-\epsilon_{jk}A_{j.k}} d\epsilon_{jk} \\
    &= \prod_{k} \frac{\Gamma(\mu_2/\gamma_k+d_{j.k}-l)}{\Gamma(\mu_2/\gamma_k)} \frac{(1/\gamma_k)^{\mu_2/\gamma_k}}{(1/\gamma_k+A_{j.k})^{(\mu_2/\gamma_k+d_{j.k}-l)}}
\end{align*}
%
where: $A_{ijk} = e_{ijk}\text{ e}^{(\boldsymbol{\beta} ^T \boldsymbol{x}_{ij} + \phi_k)}$, $A_{j.k} = \sum_{i}A_{ijk}$ and $A_{j..} = \sum_{i,k} A_{ijk}$.\\\\
Gathering everything together:
\begin{align*}
    L_j &= \prod_{i,k}[\text{e}^{(X_{ij}\beta + \phi_k)}]^{d_{ijk}} \prod_{k}\sum_{l=0}^{d_{j.k}}\binom{d_{j.k}}{l} \frac{\Gamma(\mu_1/\nu+l)}{\Gamma(\mu_1/\nu)} \frac{(1/\nu)^{\mu_1/\nu}}{(1/\nu+A_{j..})^{(\mu_1/\nu+l)}} \cdot \\
    & \cdot \frac{\Gamma(\mu_2/\gamma_k+d_{j.k}-l)}{\Gamma(\mu_2/\gamma_k)} \frac{(1/\gamma_k)^{\mu_2/\gamma_k}}{(1/\gamma_k+A_{j.k})^{(\mu_2/\gamma_k+d_{j.k}-l)}}
\end{align*}
The full likelihood is $L = \prod_{j=1}^{N} L_j$ and then we derive the full log-likelihood simply computing 
$l = \text{log}L = \sum_{j=1}^{N} \text{log}L_j$. Finally:\\
\begin{align*}
    l &= \sum_{j=1}^{N} \left[ \sum_{i,k} d_{ijk}(\boldsymbol{\beta}^T \boldsymbol{x}_{ij} +\phi_k) - \frac{\mu_1}{\nu}\text{log}(1+\nu A_{j..}) + \sum_{k} \left[ \frac{-\mu_2}{\gamma_k} \text{log}(1+\gamma_kA_{j.k})\right]\right] \\
    & \quad + \sum_{j=1}^{N} \left[ \sum_{k}  \text{log}\left( \sum_{l=0}^{d_{j.k}}\binom{d_{j.k}}{l} \frac{\Gamma(\mu_2/\gamma_k+d_{j.k}-l)}{\Gamma(\mu_2/\gamma_k)} \frac{\Gamma(\mu_1/\nu+l)}{\Gamma(\mu_1/\nu)} \frac{(A_{j.k}+1/\gamma_k)^{(l-d_{j.k})}}{(A_{j..}+1/\nu)^l} \right)  \right]
\end{align*}
%
%

\section{Proof of posterior frailty estimates and posterior frailty variance}\label{proof_posterior_frailty_estimates_posterior_frailty_variance}
As we anticipated in Section~\ref{adapted_paik_eam}, no suggestions are provided in \cite{wintrebert2004centre} to indicate how to compute any posterior frailty estimate. We decide to repeat the whole procedure made for the \textit{Time-Invariant Shared Gamma-Frailty Model} and to derive this estimate as the \textit{empirical Bayes estimate}, which correspond to the expected value of the frailty distribution given the event history. This decision is based upon the fact that both models have a Gamma as frailty distribution and similar structure of the hazard function, if isolating the frailty term.\\\\
Following \cite{balan2020tutorial}, we need first to define the \textit{Laplace transform} of the gamma density, with parameters $\zeta, \psi$, and then derive its generic $k$th derivative, as:
\begin{align*}
    \mathcal{L}(c) &= \left( \frac{\psi}{\psi + c} \right)^{\zeta} \\
    \mathcal{L}^{(k+1)}(c) &= - \mathcal{L}^{(k)}(c) \cdot \frac{(\zeta + k)}{(\psi + c)}
\end{align*}
At this point, both the Laplace transform of the frailty $Z$, given the event history $F(\tau)$ up to time instant $\tau$, and its expectation are given by:
\begin{align*}
    \mathcal{L}_{Z|F(\tau)}(c) &= \frac{\mathcal{L}^{(N(\tau))}(c + H_{\bullet}(\tau))}{\mathcal{L}^{(N(\tau))}(H_{\bullet}(\tau))} \\
    E[Z|F(\tau)] &= - \frac{\mathcal{L}^{(N(\tau)+1)}(H_{\bullet}(\tau))}{\mathcal{L}^{N(\tau)}(H_{\bullet}(\tau))} \quad \quad c=0
\end{align*}
from which we derive the frailty estimate for group $j$ up to $\tau$, as:
\begin{equation*}
    \hat{Z}_j(\tau) = E[Z_j|F_j(\tau)] = \frac{\hat{\zeta} + N_j(\tau)}{\hat{\psi} + H_{j,\bullet}(\tau)} 
\end{equation*}
and the frailty variance, always for group $j$ up to $\tau$, as:
\begin{equation*}
    Var[Z|F_j(\tau)] = \frac{\hat{\zeta} + N_j(\tau)}{(\hat{\psi} + H_{j,\bullet}(\tau))^2}
\end{equation*}
More in detail, our proposals for $\hat{\alpha}_j$ and $\hat{\epsilon}_{jk}$ are:
\begin{equation}\label{empirical_bayes_estimates}
    \hat{\alpha}_j = \frac{(\hat{\mu}_1/\hat{\nu}) + N_j}{(1/\hat{\nu}) + \hat{H}_{j,\bullet}} \quad \quad 
    \hat{\epsilon}_{jk} = \frac{(\hat{\mu}_2/\hat{\gamma_k}) + N_j(I_k)}{(1/\hat{\gamma_k}) + \hat{H}_{j,\bullet}(I_k)} \quad \quad \hat{Z}_{jk} = \hat{\alpha}_j + \hat{\epsilon}_{jk} \quad \forall j,k
\end{equation}
and their variance is:
\begin{align}
    Var(\hat{\alpha}_j) &= \frac{(\hat{\mu}_1/\hat{\nu}) + N_j}{((1/\hat{\nu}) + \hat{H}_{j,\bullet})^2} \quad \quad 
    Var(\hat{\epsilon}_{jk}) = \frac{(\hat{\mu}_2/\hat{\gamma_k}) + N_j(I_k)}{((1/\hat{\gamma_k}) + \hat{H}_{j,\bullet}(I_k))^2} \label{posterior_variance_terms}\\
    Var(\hat{Z}_{jk}) &= Var(\hat{\alpha}_j) + Var(\hat{\epsilon}_{jk}) \quad \quad  \forall j,k \nonumber
\end{align}
with:
\begin{equation}\label{cumulative_hazard_posterior}
    \hat{H}_{j,\bullet}(I_k) = \sum_i \int_{I_k} Y_{ij}(s) h_i(s) ds = \sum_i e_{ijk} Y_{ij}(I_k) h_i(I_k) 
\end{equation}
To compute the cumulative hazard function $\hat{H}_{j,\bullet}(I_k)$ for each group $j$ and interval $I_k$, we have to solve the temporal integral through the variable $e_{ijk}$. Then define the at-risk indicator $Y_{ij}(I_k)$, which assumes value $1$ if unit $i$ is at-risk in this interval (i.e. he/she has not faced the event yet) and $0$ otherwise (i.e he/she failed), and then counts the number of events happened in each group and interval to obtain $N_j(I_k)$. Concerning the hazard function $h_i(I_k)$, it must not contain the frailty term and can be individually computed as: $h_i(I_k) = exp(\boldsymbol{\beta}^T \boldsymbol{x}_{ij}  + \phi_k)$. Once all these quantities are known, it is possible to derive the posterior estimates for $\epsilon_{jk}$.\\ On the other hand, since $\alpha_j$ does not depend on time, we produce a unique posterior estimate for each group and both $N_j$ and $\hat{H}_{j, \bullet}$ must be evaluated at the end of the follow-up, which also implies summing all the interval-result $\hat{H}_{j,\bullet}(I_k)$ $\forall k$ to compute $\hat{H}_{j, \bullet}$.\\\\
However, the procedure applied so far does not take into consideration the fact that $E[Z_{jk}] = \mu_1 + \mu_2 = 1$. One possibility to solve a posteriori this problem consists in individuating the maximum value assumed by both terms $\hat{\alpha}_j$ and $\hat{\epsilon}_{jk}$ of (\ref{empirical_bayes_estimates}) (i.e. $\hat{\alpha}_{max}$ and $\hat{\epsilon}_{max}$) and then dividing the estimated terms by the correspondent maximum value, as follows: $\hat{\alpha}_j = \hat{\alpha}_j / \hat{\alpha}_{max}$, $\hat{\epsilon}_{jk} = \hat{\epsilon}_{jk} / \hat{\epsilon}_{max}$. \\
Similarly, also the posterior variances needs to be adjusted considering the constraint on the frailty mean. Therefore, we proceed as follows to obtain the final form of the posterior frailty variance:
\begin{align*}
    & Var(\hat{\alpha}_j/\hat{\alpha}_{max}) = \frac{(\hat{\mu}_1/\hat{\nu}) + N_j}{((1/\hat{\nu}) + \hat{H}_{j,\bullet})^2} \cdot \frac{1}{(\alpha_{max})^2} \\
    & Var(\hat{\epsilon}_{jk}/\hat{\epsilon}_{max}) = \frac{(\hat{\mu}_2/\hat{\gamma_k}) + N_j(I_k)}{((1/\hat{\gamma_k}) + \hat{H}_{j,\bullet}(I_k))^2} \cdot \frac{1}{(\epsilon_{max})^2} \\
    & Var(\hat{Z}_{jk}) = Var(\hat{\alpha}_j/\hat{\alpha}_{max}) + Var(\hat{\epsilon}_{jk}/\hat{\epsilon}_{max}) \quad \quad  \forall j,k 
\end{align*}
%
%

\section{Time-Independent Shared Frailty Cox Model}
\label{app:example_time_unvarying}

To fit a classical Cox proportional hazards model, we first need to define the event status, which can be derived as described in Section \ref{inputs}. Specifically, we can define the status variable as follows:

\begin{CodeChunk}
\begin{CodeOutput}
R> data("data_dropout")
R> data_dropout$status = ifelse(data_dropout$time_to_event < 6.1, 1, 0)
R> data_dropout$time_to_event = as.numeric(data_dropout$time_to_event)
\end{CodeOutput}
\end{CodeChunk}

Moreover, we can make use of \texttt{frailtyPenal} in \pkg{frailtypack} \citep{frailtypack}

\begin{CodeChunk}
\begin{CodeOutput}
R> library(frailtypack)
R> frailty_model <- frailtyPenal(Surv(time_to_event, status) ~ cluster(group) + 
                              Gender + CFUP, data = data_dropout, 
                              cross.validation = FALSE,
                              n.knots = 20, kappa = 1, hazard="Splines")
R> frailty_model$coef
   GenderMale       CFUP 
    0.3297152 -1.2953989 
R> frailty_model$beta_p.value
   GenderMale        CFUP 
  0.002788997 0.000000000 
\end{CodeOutput}
\end{CodeChunk}


The baseline hazard function, which reflects the instantaneous risk of dropout over time, can be visualized after fitting the model. To plot the estimated baseline hazard, we can apply a smoothing spline and normalize the hazard function by its area, as shown below

\begin{CodeChunk}
\begin{CodeOutput}
frailty_time <- frailty_model$x
frailty_hazard <- frailty_model$lam[,1,1]

smoothingSpline = smooth.spline(frailty_time, frailty_hazard, spar=0.35)
area = 0 
for(ii in 1:(length(smoothingSpline$y)-1)){
  area<-area+(smoothingSpline$x[ii+1]-smoothingSpline$x[ii])*smoothingSpline$y[ii+1]
}
smoothingSpline$y <- smoothingSpline$y/area

plot(smoothingSpline$x[17:98], smoothingSpline$y[17:98], type = 'l', col="black",
     main = "Estimated baseline hazard function", 
     xlab = "Time [semesters]", ylab="Instantaneous risk of failure")
\end{CodeOutput}
\end{CodeChunk}

\begin{figure}[]
    \centering
    \includegraphics[width=0.75\textwidth]{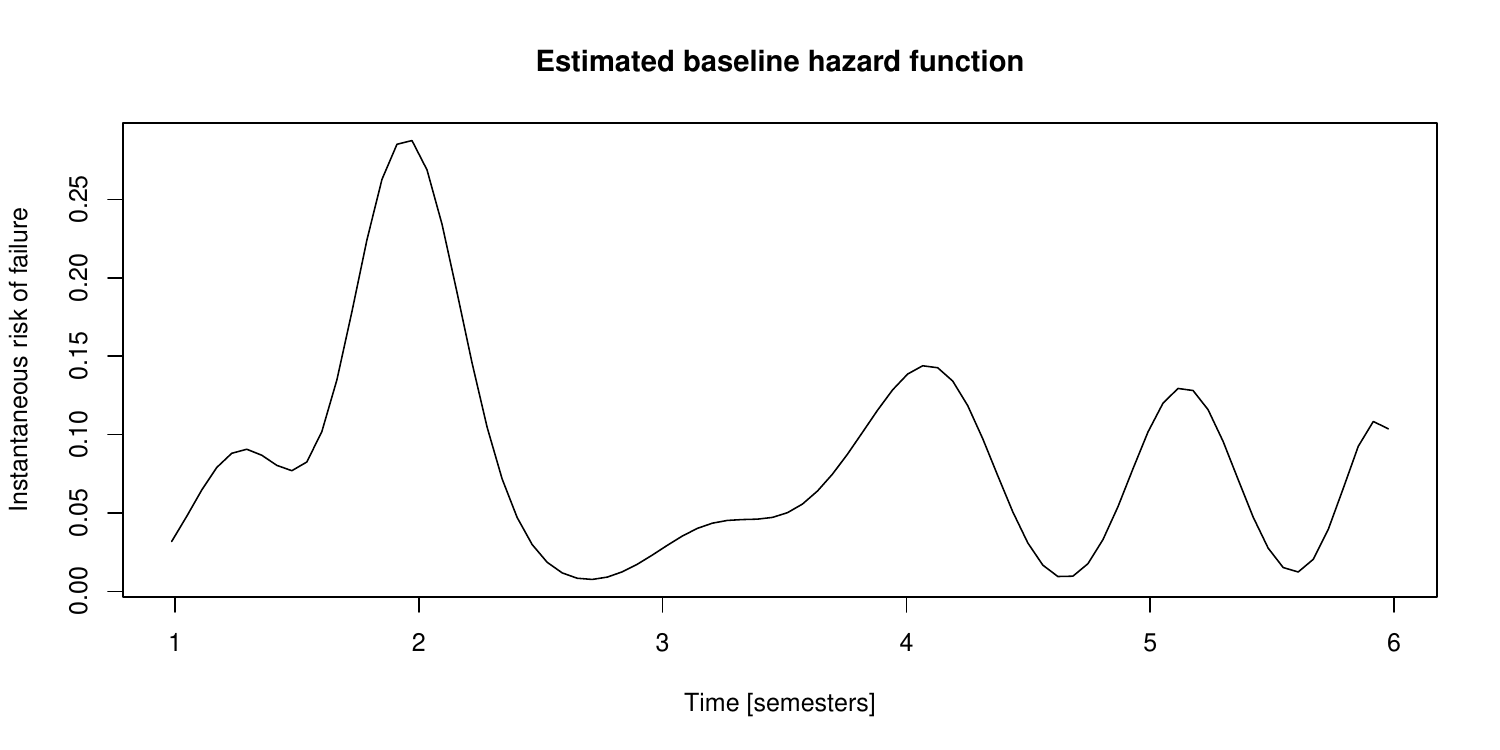}
    \caption{Estimated baseline hazard function through a Time-Independent Shared Frailty Cox Model.}
    \label{baseline_hazard}
\end{figure}

The resulting baseline hazard function is shown in Figure \ref{baseline_hazard}. 

\end{appendix}

\bibliographystyle{chicago} 
\bibliography{Biblio}

\end{document}